\documentstyle[twocolumn,aps,epsfig,floats]{revtex}
\begin{document}

\title{Short-Time Effects on Eigenstate Structure in Sinai Billiards and
Related Systems}
\author{L. Kaplan\thanks{kaplan@tornado.harvard.edu}
 \\Department of Physics and Society of Fellows,\\ Harvard
University, Cambridge, MA 02138\\ \vskip 0.1in
E. J. Heller\thanks{heller@physics.harvard.edu}
 \\ Department of
Physics and Department of Chemistry,\\ Harvard University, Cambridge, MA 02138}
\maketitle

\begin{abstract}
There is much latitude between the requirements of Schnirelman's
theorem regarding the ergodicity 
of individual high-energy eigenstates of 
classically chaotic systems on the 
one hand, and the extreme requirements  of random matrix theory on the 
other.  It seems likely that some eigenstate statistics  and long-time
transport behavior bear  nonrandom   imprints
of the underlying classical dynamics while simultaneously
obeying Schnirelman's theorem.  Indeed this was shown earlier in the 
case of systems which approach classical ergodicity slowly, and is also 
realized in the scarring of eigenstates, even in the
$\hbar \to 0$ limit, along unstable 
periodic orbits and their manifolds.  Here we demonstrate the nonrandom 
character of eigenstates of Sinai-like systems.  We show that 
mixing between channels in Sinai systems is dramatically 
deficient compared to   random matrix theory predictions.
The deficit {\it increases} as $|\log \hbar|$ for $\hbar\to 0$,
and is due to the vicinity of the measure zero set of orbits which never 
collide with the Sinai obstruction.
Coarse graining to macroscopic  scales recovers the Schnirelman
result.  Three systems are investigated here: a Sinai-type billiard,
a quantum map which possesses the essential properties of the
Sinai billiard,
and a unitary map corresponding to a quasirandom Hamiltonian.
Various wavefunction and long-time transport
statistics are defined, theoretically investigated,
and compared to numerical data.
\end{abstract}

\section{Introduction}

In recent years, much attention has been paid to the structure of
quantum eigenstates in systems with a chaotic or ergodic classical
analogue. For integrable systems, EBK quantization provides an intuitive
understanding of classical--quantum correspondence, associating quantum
wavefunctions with the invariant tori of the underlying classical dynamics.
In a classically ergodic system, the typical trajectory fills an entire energy
hypersurface at long times, and it is natural to conjecture that a typical
high-energy eigenstate of such a system similarly has intensity distributed
evenly over an entire energy shell. Thus, Berry suggested in 1983 that an
eigenstate of a classically ergodic system should look locally like a random
superposition of plane waves of fixed energy, with momenta pointing in all
possible directions~\cite{berry}. Similarly, Bohigas, Giannoni, and
Schmit~\cite{bgs} proposed that the quantum
properties of a classically chaotic system should correspond to those of
random matrix theory (RMT). This implies that wavefunction intensity should
be distributed over an entire energy surface, with the wavefunction
amplitudes at distant points behaving as independent Gaussian variables.

The conjecture that chaotic eigenstates obey RMT statistics is a statement
about quantum structure at the scale of a single wavelength in position space
(or on the
scale of a single
channel in momentum space, or more generally, on a mesh of cell
size O($\hbar$) in phase
space). Rigorous results on quantum ergodicity, however, mostly address 
structure on classically large scales, in the limit where $\hbar$ becomes
small compared to the phase-space region over which wavefunction intensity
is being smoothed. Specifically, theorems by Schnirelman, Zelditch,
and Colin de Verdiere (SZCdV)~\cite{szcdv}
state that for a classically defined operator, the
expectation value over almost all wavefunctions converges to the
microcanonical average of the classical version of the operator, in
the $\hbar \to 0$ limit. Since the classical symbol of the operator is kept
fixed as the limit is taken, these theorems provide information only
about the coarse-grained structure of the eigenstates, and not about the 
structure at quantum mechanical scales.

Wavefunction scarring, the anomalous enhancement (or suppression) of
intensity near an unstable periodic orbit, is a well-known
example of non-RMT behavior of eigenstates in a classically chaotic system.
The distribution of wavefunction intensities on a fixed periodic orbit
can be computed in the semiclassical limit
using the linear and nonlinear theory of scars~\cite{scars,nlscar,sscar},
and is found to be very different from
the Porter-Thomas prediction of RMT. Furthermore, upon ensemble
averaging, a power-law wavefunction intensity distribution
tail is obtained (and numerically
observed) in chaotic systems,
in contrast with the exponential falloff prediction of RMT.
The fraction of strongly scarred states remains finite in the $\hbar \to 0$
limit. Nevertheless, scarring poses no threat to the SZCdV ergodicity
condition, because the size of the scarred phase-space region
surrounding the orbit
scales as $\hbar$, tending to zero in the semiclassical
limit. A finite intensity enhancement factor
affecting an ever smaller region of phase space is entirely consistent
with ergodicity on coarse-grained scales. However, the scarring phenomenon
does have very significant effects on physical quantities that depend
on fine-scale structure, such as conductances and decay rates through small
(or tunneling) leads~\cite{decayscar2,decayscar}.

Another example of markedly non-RMT behavior still consistent with
SZCdV coarse-grained ergodicity is found in the ``slow ergodic" systems,
such as the tilted wall billiard and the sawtooth potential
kicked map~\cite{wqe}.
In these systems, the classical rate of exploration in momentum
space is logarithmically slow, and for large  $\hbar^{-1}$,
the number of channels
occupied by a typical eigenstate scales only as $\hbar^{-1/2}\log\hbar^{-1}$,
constituting an ever
decreasing fraction of the $O(\hbar^{-1})$ total available number of channels.
However, the ``bright"
channels occupied by a given
wavefunction tend to be evenly distributed over the entire phase space,
and thus coarse-grained ergodicity still holds in the limit,
even though the wavefunctions
are becoming less and less ergodic at the single channel scale as
$\hbar \to 0$. The present paper extends the indications  of
non-RMT ``clumping'' of wavefunction density beyond the effects of scarring.
Moreover, we  use as our examples
the original paradigm of
classical Hamiltonian chaos, the Sinai billiard,
and closely related quantum maps. 

The remainder of this paper is organized as follows: in the next section
we discuss measures of ``microscopic" ($\hbar-$scale) quantum ergodicity,
including various inverse participation ratios and channel-to-channel
transport measures. Then in Section~\ref{stdyn} the connection is made 
between these stationary properties and the short-time dynamics of a
quantum system. In Section~\ref{skm} the Sinai kicked map, a one-dimensional
model for the Sinai billiard is introduced and discussed. Strong
deviations from single-channel quantum ergodicity are predicted, and
distributions for various quantities are computed, that differ greatly
from RMT expectations. We see that classical methods can be used to determine
the non-ergodic structure of the quantum wavefunctions, even though
the classical dynamics is entirely ergodic.
Quantitative comparison with numerical data follows in
Section~\ref{numtests}. In Section~\ref{billiard} a similar analysis
follows for the two-dimensional Sinai billiard system, a paradigm
of classical and quantum chaos. Here again
strongly non-RMT wavefunction intensity distributions are
predicted and observed. In Section~\ref{matrix} a simple matrix model
is presented and studied, some of the statistical properties of which
correspond to those of the Sinai systems. Similarities and differences
between the Sinai systems and the matrix ensembles are discussed.
In the final section we sum up the results and
discuss certain directions for the future.

\section{Measures of $\hbar$-scale ergodicity}
\label{measures}

We now review 
some important concepts related to the quantitative
measurement of quantum structure and transport at ``microscopic" (i.e.
single-wavelength or single-channel) scales.
An alternative discussion may be found in~\cite{wqe}.

Consider a classically ergodic system with quantum eigenstates $|\xi\rangle$
and a test state basis $|a\rangle$. The test basis can be chosen to be
the set 
of position states, momentum states, phase-space Gaussians, or any other
set of states motivated by the physics of the problem. Often the test basis
will be taken to be the set of eigenstates of a zeroth-order Hamiltonian
$H_0$, of which the full system Hamiltonian $H=H_0+\delta H$ is a
perturbation. One is then interested in determining whether the true
eigenstates $|\xi\rangle$ have a nontrivial structure in the states
$|a\rangle$,
or whether the perturbation $\delta H$ is sufficiently large so as to 
completely randomize the matrix elements $\langle a |\xi \rangle$. Thus,
in the case of tight-binding models (e.g. Anderson localization),
one may consider $H_0$ to be the Hamiltonian with on-site energies only,
and $|a\rangle$ to be the position states. The matrix
elements $\langle a |\xi \rangle$ then measure the degree of
localization in position space
as the hopping matrix elements are turned on. Similarly, in a scattering
problem one often finds it useful to use momentum states or channels as
the reference basis $|a\rangle$, and look for localization of the full
eigenstates relative to this basis.

For simplicity, we assume that the classical dynamics given by $H$
completely mixes
the states $|a\rangle$ with each other,
so that no conservation laws prevent each of  the eigenstates $|\xi\rangle$
from having equal overlaps with all of the test states. In the
presence of energy conservation or other conserved quantities, the
formalism outlined below needs to be modified to take into account
constraints imposed by the classical symmetries. This can be done in a
straightforward way by, for example, taking the test states $|a\rangle$
to be coherent states (Gaussians) in phase space. Then it is easy to compute
the classical intersection of each such Gaussian with any given energy
hypersurface, and the actual quantum intensities $|\langle  a |\xi \rangle|^2$
can be normalized by this classical result. In this way one can easily
identify the
degree of eigenstate localization (or deviation from ergodicity) due
to quantum effects, as opposed to purely classical constraints.
See Ref.~\cite{clasconstr} for a fuller discussion.

We will then focus on the set of (properly normalized) overlap intensities
\begin{equation}
P_{a\xi} = |\langle  a |\xi \rangle|^2
\end{equation}
to devise measures of ``microscopic"
localization or ergodicity in the system under study. In RMT
(a natural baseline assumption in the absence of dynamical information
about our system), the $\langle  a |\xi \rangle$
are predicted to be given by uncorrelated
random Gaussian variables, real or complex. The intensities
$P_{a\xi}$ then follow a $\chi^2$ distribution, of one or two degrees of
freedom, respectively. Quantum localization will produce an excess of
very large and very small intensities, compared to this baseline result.
For convenience, we adopt the normalization where the mean intensity
is set to unity:
\begin{equation}
\langle P_{a\xi} \rangle _a = \langle P_{a\xi}\rangle_\xi =1 \,.
\end{equation}
Here the averages $\langle \ldots\rangle_\xi$
 are taken over all eigenstates $|\xi\rangle$:
\begin{equation}
\langle P_{a\xi}\rangle_\xi = {1 \over N} \sum_{\xi=1}^N P_{a\xi} \,,
\end{equation}
where $N$ is the total number of states accessible from $|a\rangle$ (the
dimension of the effective Hilbert space). The averaging 
$\langle\ldots\rangle_a$ over basis states $|a\rangle$ is defined similarly.

It is often convenient to compress the intensity information into
the set of (local) inverse participation values (IPR's)\cite{clasconstr}:
\begin{equation}
\label{ipra}
{\rm IPR}_a = P_{aa} = \langle P_{a\xi}^2\rangle_\xi \,.
\end{equation}
$P_{aa}$  is a convenient alternative notation for ${\rm IPR}_a$, as we
will see below when we discuss transport in Eqs.~\ref{autocorr} to
\ref{pabdyn}.
For a given
test state $|a\rangle$, the IPR at $|a\rangle$
gives the first non-trivial moment
of the $P_{a\xi}$ distribution (namely, the ratio of the mean squared
intensity to the square of the mean),
and thus gives a concise measure of the degree
of localization at $|a\rangle$~\footnote{A slightly different measure of
eigenstate localization at a given test state,
defined in analogy with classical entropy ideas, is
discussed in~\cite{mirbach}.}.
The IPR measures the inverse of the
fraction of eigenstates which have significant intensity at $|a\rangle$.
Thus, equal
intensities of all the eigenstates at $|a\rangle$ would imply
${\rm IPR}_a =1$; this level of ergodicity is of course almost never
achieved in a chaotic system. Gaussian random fluctuations (RMT) produce
IPR's of 3 (for real overlaps $\langle a|\xi\rangle$)
or 2 (for complex overlaps). IPR's exceeding
the appropriate baseline value signal the presence
of a localization mechanism beyond RMT.
In the extreme localization limit where one eigenstate has all its intensity
at $|a\rangle$, we obtain the maximum possible value,
${\rm IPR}_a =N$. [A prime example of such extreme behavior is the case
of the ``bouncing ball" states~\cite{bb}, associated with non-isolated, 
marginally stable classical periodic motion. Such classical trajectories
can trap a quantum wavepacket $|a\rangle$
for a time comparable to (or even longer than)
the time at which individual quantum states are resolved, causing the
wavepacket to have $O(1)$ overlap with only one or a few eigenstates,
and leading to ${\rm IPR}_a=O(N)$. From the SZCdV theorems, we easily
see that that the fraction of bouncing ball states must tend to
zero in the $\hbar \to 0$ limit. This kind of localization is easily
visible to the naked eye; other kinds of localization, where the number of
eigenstates having intensity at the test state $|a\rangle$
is large compared to
$1$ but small compared to the total number of states $N$, may be less
easy to detect visually
but may also be a statistically more important correction
to RMT predictions, surviving at arbitrarily small values of $\hbar$.]

One can similarly define an eigenstate-specific IPR in the
$|\xi\rangle$-basis:
\begin{equation}
\label{pnndef}
{\rm IPR}_\xi = P_{\xi\xi} = \langle P_{a\xi}^2\rangle_a \,,
\end{equation}
where the average is taken over the test basis $|a\rangle$. This of course
measures the inverse of the fraction of phase space occupied by a given
eigenstate $|\xi\rangle$,
in the $|a\rangle$-basis. A global IPR can also be defined:
\begin{equation}
\label{globipr}
{\rm IPR} =  \langle{\rm IPR}_a\rangle_a = \langle{\rm IPR}_\xi\rangle_\xi\,.
\end{equation}
This last quantity measures the inverse fraction
of phase space occupied by the
{\it average} eigenstate (or equivalently, the inverse fraction of eigenstates
that have intensity at an average location),
and can serve as a simple figure of merit for the degree of quantum
localization in a given system.

We can relate eigenstate localization to dynamics in the following way.
Let 
\begin{equation}
\label{autocorr}
A_{aa}(t) = \langle a| e^{-i Ht}|a \rangle
\end{equation}
be the return amplitude for state $|a\rangle$ to come back to itself after
time $t$. Given a non-degenerate spectrum, the mean return probability
of the state $|a\rangle$ at long times is proportional to its eigenstate IPR:
\begin{equation}
P_{aa} = N \langle|A_{aa}(t)|^2\rangle_t \,,
\label{paadef}
\end{equation}
as is easily seen by inserting complete sets of eigenstates on the right hand
side.
Here the time average on the right hand side
of Eq.~\ref{paadef}
is taken over times large compared to the Heisenberg
time $T_H$, i.e. $\hbar$ over
the mean level spacing.

Similarly, we can relate long-time transport to eigenstate correlations.
Defining the transport amplitude
\begin{equation}
A_{ab}(t) = \langle a| e^{-i Ht}|b \rangle\,,
\end{equation}
and a cross-correlation analogous to $P_{aa}$,
\begin{equation}
\label{pabdef}
P_{ab} = \langle P_{a\xi}P_{b\xi}\rangle_\xi \,,
\end{equation}
we immediately see 
\begin{equation}
\label{pabdyn}
P_{ab} = N \langle|A_{ab}(t)|^2\rangle_t \,,
\end{equation}
where again the time average has been taken of the long-time transport
probability
from $|b\rangle$ to $|a\rangle$. Of course the total probability summed
over final states for any given initial state is normalized:
\begin{equation}
\langle P_{ab}\rangle_a =1
\end{equation}
for each $b$. The simplest non-trivial quantity
which will measure the fluctuation 
in the probabilities of being in various final states $|b\rangle$ given
an initial state $|a\rangle$ (or vice versa) is:
\begin{equation}
\label{qdef}
Q_a = \langle P_{ab}^2\rangle_b \,.
\end{equation}
Roughly speaking, $Q_a$ measures the inverse fraction of all channels
that are accessible at long time from channel $a$.
In RMT, all the transport probabilities $P_{ab}$ are equal to unity with small
fluctuations (except for the enhanced return probabilities $P_{aa}=2$ or $3$),
so $Q_a=1$ for each $|a\rangle$ in the $N \to \infty$ semiclassical limit.
$Q_a>1$ indicates uneven visiting of the available
state space starting in the initial state $|a\rangle$, and the overall
ergodicity of long-time transport can again be summarized in
\begin{equation}
Q=\langle Q_a \rangle_a \,.
\end{equation}

In the slow ergodic systems such as the tilted wall billiard and the
sawtooth potential kicked map~\cite{wqe}, a highly anomalous IPR
measure was predicted and observed for small $\hbar$,
with the system-averaged IPR (Eq.~\ref{globipr})
scaling as $\sqrt {\hbar^{-1}} /\log {\hbar^{-1}}$. Semiclassically,
the degree of localization in such systems is even stronger, with the IPR
scaling as ${\hbar^{-1}} /\log {\hbar^{-1}}$. The difference
is caused by diffraction, 
which dominates the phase space exploration and increases
by $\sqrt {\hbar^{-1}}$
the fraction of phase space occupied by a typical eigenstate. These same
diffractive effects lead to almost perfect long-time transport between
channels, with the transport measure $Q$ an $\hbar$-independent constant
in the $\hbar \to 0$ limit.  

\section{Short-time dynamics and the eigenstates}
\label{stdyn}

We now discuss the way in which short-time dynamics produces
lasting effects on stationary properties, such as eigenstate localization
and long-time transport\cite{scars,nlscar,clasconstr}.
Define the local density of states
(LDOS) at $|a\rangle$ as the Fourier transform of the autocorrelation 
function $A_{aa}(t)$:
\begin{equation}
\label{ldos}
S_a(E)= {N \over 2\pi}
\int\limits_{-\infty}^\infty dt \;
e^{iEt}A_{aa}(t) = \sum_\xi P_{a\xi} \delta(E-E_\xi)\,.
\end{equation}
The linearity of the Fourier transform implies that large short-time
recurrences
in $A_{aa}(t)$ get ``burned into" the spectrum, producing an envelope which
must be the smoothed version of the full spectrum $S_a(E)$. Thus, let
\begin{equation}
A_{aa}(t) = A_{aa}^{\rm short}(t) + A_{aa}^{\rm long}(t).
\end{equation}
(The most convenient separation
between short- and long-time recurrences
is situation-dependent, as we will see below.) Then the full spectrum
is given by the sum of a short-time
envelope and a high-frequency oscillatory structure,
coming from the long-time dynamics, that is superimposed on top of that
envelope:
\begin{equation}
S_a(E) = S_a^{\rm short}(E) + S_a^{\rm long}(E).
\end{equation}

In the presence of chaos, the number of classical returning
trajectories leading
from any state $|a\rangle$ back to itself
grows exponentially with time (with some Lyapunov exponent
$\lambda$). It is then convenient to classify as ``short time" those returns
that are governed by one or a small number of classical paths, and
as ``long time" those that arise from interference between many classical
paths, and for which a statistical description is valid.
Due to the exponential
proliferation, the dividing line between these regimes is sharply defined
in the semiclassical limit $\hbar \to 0$, being given by the mixing time
$T_{\rm mix} \sim {1 \over \lambda} \log \hbar^{-1}$. Of course, no harm would
be done were we to
err on the safe side by computing explicitly intermediate-time
return amplitudes
which could instead have been treated statistically.

Now, in the presence of chaos, the long-time returning amplitude at time
$t$ is expected
to fill the initial test state $|a\rangle$ in a uniform, unbiased
way~\cite{nlscar}, so that the subsequent evolution of this newly
returned amplitude is equivalent to the evolution of the original state.
More explicitly, we may write for small $\tau$
\begin{eqnarray}
A_{aa}^{\rm long}(t+\tau)&=&\langle a|a(t+\tau)\rangle \nonumber \\ &=&  
\langle a|a(t)\rangle \langle a(t) |a(t+\tau)\rangle + \cdots \nonumber
\\ & = & A_{aa}^{\rm long}(t)A_{aa}^{\rm short}(\tau)+ \cdots \,.
\end{eqnarray}
Thus, randomly returning amplitude at long time $t$ leaves its imprint
on nearby times $t+\tau$, and following \cite{nlscar} we may write the
full long-time amplitude to return as a convolution~\footnote{Discrete-time
notation is used here for simplicity and because it is most useful
for the repeated scattering situation discussed below.
Refer to Ref.~\cite{nlscar} for a full
treatment of continuous time, which involves introducing an
additional very short time scale associated with the initial decay
of the wavepacket $|a\rangle$ (or with $\hbar$ divided by the energy
uncertainty of $|a\rangle$).}
\begin{equation}
\label{convol}
A_{aa}^{\rm long}(t)=\sum_\tau A_{aa}^{\rm rnd}(t-\tau)A_{aa}^{\rm short}
(\tau)\,,
\end{equation}
where $A_{aa}^{\rm rnd}$ is an uncorrelated  Gaussian random variable;
\begin{equation}
\langle A_{aa}^{{\rm rnd} \ast}(t+\Delta)A_{aa}^{\rm rnd}(t)\rangle=
 {1 \over N} 
\delta_{\Delta 0} \,,
\end{equation}
the averaging being performed over long times $t$ or over an appropriate
ensemble. The $1/N$ factor provides the right normalization for the 
probability to return in the absence of nontrivial short-time overlaps,
i.e. when $A_{aa}^{\rm short}(\tau)=\delta_{\tau 0}$.

Fourier transforming, the convolution in Eq.~\ref{convol} leads
to multiplication of the smooth short-time envelope by random
oscillations in the energy domain:
\begin{equation}
S_a^{\rm long}(E)=S_a^{\rm short}(E)S_a^{\rm rnd}(E) \,.
\end{equation}
At very long times (beyond the Heisenberg time $T_H$, which scales
as $\hbar$ divided by the mean level spacing), the spectrum
$S_a(E)$ becomes resolved into individual spectral lines
\begin{equation}
\label{linespec}
S_a(E)=\sum_\xi r_{a\xi} S_a^{\rm short}(E) \delta(E-E_\xi) \,,
\end{equation}
where $r_{a\xi}$ has the statistical properties of a $\chi^2$ variable.

That is, 
\begin{equation}
\langle r_{a\xi}\rangle=1 \;\;\;\;\; \langle r_{a\xi}^2 \rangle=F \,,
\end{equation}
where averaging may be performed over eigenstates $|\xi\rangle$,
test states $|a \rangle$, or over an ensemble, and the constant $F$
is given by $2$ or $3$, for complex or real eigenstates, respectively.
RMT predictions are recovered in the dynamics-free
limit $S_a^{\rm short}(E)=1$; short-time recurrences cause
$S_a^{\rm short}(E)$ to vary with energy. This variation in turn leads
to larger-than-expected wavefunction intensities $P_{a\xi}$ at some
energies and smaller intensities at others, corresponding to an enhanced
IPR and deviation from microscopic ergodicity.

The formalism outlined above has already been used to study the statistical
properties of the scarring phenomenon, the anomalous enhancement of
certain quantum eigenstates along the unstable periodic orbits of the
corresponding classical chaotic system. There, the test state
$|a\rangle$ is a wavepacket launched on or near the classical periodic
orbit, and short-time quantum recurrences can be computed analytically
(for small $\hbar$) in terms of the monodromy matrix and action
of the classical orbit. One finds that the IPR for a test state on the orbit
scales inversely with the instability exponent $\lambda$ of the
orbit~\cite{nlscar}, and the full distribution of wavefunction intensities
on and off the orbit can be computed as a function of $\lambda$~\cite{sscar}.
Averaging over an ensemble of chaotic systems was shown to produce
a power-law tail in the intensity distribution, dominated by very strongly
scarred states, and in contrast to the exponential tail prediction of RMT.
Antiscarred states (ones with anomalously low intensity in certain regions
of phase space) are of great importance in open systems: for example,
they have been shown to dominate the long-time quantum
probability to remain in a classically chaotic system coupled
to the outside via a single-channel lead. The size of the effect is
exponentially large for small $\lambda$\cite{decayscar}. More recently,
optimal test states for measuring scarring (``universal scarmometers")
have been developed, which take into account an entire classical orbit
and the linearized classical dynamics in its vicinity~\cite{is}:
they provide larger IPR's and more evidence of wavefunction localization
than do simple Gaussian wavepackets.

\section{Sinai kicked maps}
\label{skm}

\subsection{Definition of system and motivation}

The Sinai billiard~\cite{sinairef}
is a prototypical example of strong classical chaos: it consists
of a point particle bouncing freely
in a rectangular cavity with hard walls, with a hard disk obstruction
placed in the center of the rectangle.
The system has positive
entropy classically for any disk size; of course, this fact becomes relevant
to the quantum mechanics 
only in the limit where the quantum wavelength is small compared to the
size of the disk. (The mixing time $T_{\rm mix}$ after which a typical
wavepacket spreads over the entire available phase space
is then short compared to the
Heisenberg time
$T_H$, defined as $\hbar$ over the mean level spacing,
at which the quantum dynamics becomes quasi-periodic and individual eigenstates
and eigenvalues begin to be resolved.)

The statistics of energy levels in the (desymmetrized)
Sinai billiard has been
found to be in good agreement with the GOE predictions of random matrix
theory~\cite{sinaigoe}.
On the other hand, the eigenstate structure of the Sinai
billiard turns out to be very different from RMT expectations, and the
inclusion of short-time dynamical effects is essential for understanding its
quantum ergodic properties. We will return to a detailed discussion of the
classical and quantum Sinai billiard in Section~\ref{billiard}.

Here we begin
with a simplified one-dimensional model
which contains most of the important structure of the original
two-dimensional system. We notice first that finding eigenstates of a given 
symmetry class in the Sinai billiard is equivalent to finding the eigenstates
in a rectangle one-fourth the original size, with a quarter-circular bump
in one of the corners (and possibly with Neumann boundary conditions along one
or both of the sides meeting at that corner). We can then
imagine finding the eigenstates using an S-matrix approach~\cite{smatrix,smil},
where one considers the scattering of channels of the ``free" rectangular
system off the quarter-circular bump. The S-matrix has a strong diagonal
component due to the straight part of the wall containing the bump, and
a complicated off-diagonal structure due to actual
scattering off the bump.
The long-time dynamics and stationary properties of the system (e.g.
eigenstates and eigenvalues) are obtained by iterating the scattering process.

From the surface of section method, we know that Hamiltonian dynamics
in a two-dimensional configuration space at fixed energy is dimensionally
equivalent
to a discrete-time
mapping of a one-dimensional system, and can in fact be reduced
to such a system.
The one-dimensional model we consider in this section is the
``Sinai kicked map", defined on a two-dimensional phase space
$(q,p)\in [0,1) \times [0,1)$:
\begin{eqnarray}
\label{eqmo}
p \to \tilde p &=& p - V'(q) \; {\rm mod} \; 1 \nonumber \\
q \to \tilde q &=& q + \tilde p \; {\rm mod} \; 1 \,.
\end{eqnarray}
The equations of motion Eq.~\ref{eqmo} can be viewed as arising from
a potential that is periodic in time:
\begin{equation}
H(q,p,t) = {p^2 \over 2} + V(q) \sum_n \delta(t-n)\,.
\end{equation}
At the beginning of every time step, the particle is ``kicked" by the potential
$V$, following which the potential is turned off and free evolution takes
place for a unit time interval. The process is then iterated to obtain the
long-time behavior. The Sinai billiard's straight wall with a bump has its
analogue in the kick potential
\begin{eqnarray}
\label{kickpot}
V(q)&=&-{K \over 2f} \left[\left(q-{1 \over 2}\right)^2
 - \left({f \over 2}\right)^2\right] \;
 \; {\rm for} \; \left|q-{1 \over 2}\right| < {f \over 2} \nonumber \\
&=& 0 \;\; {\rm otherwise,}
\end{eqnarray}
with a parabolic bump (centered at $q=1/2$)
of spatial extent $f<1$. $K$ is a constant (which we will
set to be of order unity) that determines the typical impulse exerted by the
bump. The parabolic shape of the potential bump
is chosen for simplicity only; none
of the discussion below would be affected if a semicircular or other curved
potential were used instead. The key property of the repulsive
potential $V(q)$ is
the parameter $f$, which sets the fraction of an incoming wave that is
scattered classically after one iteration of the map.

The Sinai kicked map is a hard chaotic system with no stable phase
space regions (as can be seen easily by computing the Jacobian of the
iterated mapping, using the fact that $V''(q) \le 0$ everywhere~\cite{unst}).
Like the Sinai billiard (and the Bunimovich stadium), the system has
a measure zero set of marginally unstable trajectories, given for example by
$|q-{1 \over 2}|> {f \over 2}$, $p=0$. After quantization, such orbits
will give
rise to ``bouncing ball" states~\cite{bb}, which are very strongly localized
in momentum space near $p=0$. Our primary interest, however, will be not
in this measure zero set of states, but rather in the structure of the 
``typical" quantum wavefunctions, which obey SZCdV coarse-grained
ergodicity, yet have very non-uniform structure at the single channel scale.

The quantization of kicked systems of the form Eq.~\ref{eqmo} is
straightforward and well-covered in the literature~\cite{kickmap}. A value
of $\hbar$ should be chosen so that $N=1/2\pi\hbar$, the number of Planck cells
covering the toroidal classical phase space, has an integer value. Then
an $N$-dimensional position basis for the Hilbert
space is given by $q_i=(i+\epsilon_0)/N$, $i=0 \ldots N-1$.
Similarly, the momentum
space basis is given by $p_j=(j+\epsilon_1)/N$, $j=0 \ldots N-1$.
$\epsilon_{0,1}$
form a family of possible quantization conditions (they correspond to phases
associated with circling the torus in the $p$ and $q$ directions,
respectively). The two bases are related by a discrete Fourier transform.
The quantum dynamics is now defined by the unitary $N \times N$ matrix
\begin{equation}
U =
\exp{\left[-i \left({1 \over 2} \hat p^2\right)/\hbar\right]}
\cdot
\exp{\left[-i V(\hat q)/\hbar\right]}
\,,
\end{equation}
where each factor is evaluated in the appropriate basis, and an implicit
forward and backward Fourier transform has been performed.

We are now ready to study the structure of the Floquet or scattering
eigenstates of the quantum dynamics $U$, in the $p_j$ basis. We notice first
that because of the symmetry of the kick potential $V(q)$ under $q \to -q$,
the classical system has a time reversal symmetry and a parity symmetry:
\begin{eqnarray}
T:\, t &\to& -t,\, q \to -q \nonumber \\
P:\, p &\to& -p,\, q \to -q \,.
\end{eqnarray}
It will be convenient for us to choose a nonzero value for the boundary
condition parameter $\epsilon_1$, thus breaking the parity symmetry
$P$ under quantization, while maintaining the time reversal symmetry
$T$ by setting $\epsilon_0=0$. The eigenstates are then real in the
momentum basis, and the appropriate RMT baseline is ${\rm IPR}=3$ (see
discussion following Eq.~\ref{ipra}). For an asymmetric bump or kick potential,
the quantum wavefunction intensity fluctuations would be expected to follow
a $\chi^2$ distribution of two degrees of freedom under RMT, giving
rise to the baseline value ${\rm IPR}=2$. The analysis to follow is of
course completely independent of the symmetry chosen, provided that the
appropriate baseline quantum fluctuation factor $F$ is used.

\subsection{Short-time dynamics}
\label{short}
\subsubsection{Quantum factor of two}

As suggested in the preceding section, we should
begin our analysis by examining
the classical and quantum short time dynamics of the Sinai kicked map
in momentum space. 
In analogy with the quantum return amplitude $A_{nn}(t)$ of Eq.~\ref{autocorr},
let $P^{\rm cl}_{nn}(t)$ be the classical probability to remain in state
$p_n$ after $t$ iterations of the map.~\footnote{For conventional
resons we will be using the index $n$ to label the momentum states instead
of the generic index $a$ used in Section~\ref{stdyn}.}
Classically, for any incoming momentum $p_n$, a fraction
\begin{equation}
P^{\rm cl}_{nn}(1)=1-f
\end{equation}
of all particles remain in momentum $p_n$ after one kick,
while the remaining
fraction $f$ get scattered to other momentum states. Semiclassically,
there is an {\it amplitude} $1-f$ for remaining in the incoming channel,
as can easily be seen by taking the (semiclassical)  overlap of the
initial and final states.
The {\it probability} to remain unscattered after one step is then
\begin{equation}
|U_{nn}|^2=|A_{nn}(1)|^2=(1-f)^2 \,.
\end{equation}

Notice that the quantum one step survival $|U_{nn}|^2$ 
is smaller than the classical
probability $P^{\rm cl}_{nn}(1)$
for not scattering.  In analogy to ordinary scattering
in free space, one can define  a cross section for scattering
off the defect.  In the limit of small $f$, we see from the above 
analysis that the 
quantum cross section is twice as large as 
the classical. (This is in complete analogy with 
 quantum scattering theory in free space, in which   diffraction
results in a quantum cross section twice as big as the 
classical, even in the short wavelength 
limit. Essentially, the far-field  diffraction
into the shadow zone  doubles the 
quantum cross section).  Here, there is   a quantum
probability $f(1-f)$ for diffracting into {\it nearby} channels.
Classical-quantum correspondence still holds after appropriate coarse-graining
over scales large compared to $\hbar/f$ (but still small classically) in
momentum space. As mentioned earlier in our discussion of the Sinai
billiard, we are always working in the semiclassical regime $\hbar \ll f$,
where the bump size is large compared to a wavelength, though it may be small
compared to the system size $1$.

This difference between the classical and quantum probability to 
remain in the initial channel ($(1-f) \; vs. \; (1-f)^2$) survives the 
limit $\hbar\to 0$.  In this limit the fraction of diffracted 
amplitude and the number of scattering channels into
which diffraction occurs both remain constant.
For finite $\hbar$, there will of course
be an $O(\sqrt\hbar)$
correction to the quantum amplitude $A_{nn}(1)$, as it is possible for
the horizontal ($V'(q)=0$) portion of the bump to scatter an incoming channel
$p_n$ back into itself.

\subsubsection{Multistep scattering}
\label{lorenz}
We proceed to analyze the multi-step behavior of the dynamics, particularly
the probability to remain in the initial state $p_n$. In the absence of
step-to-step correlation, the classical
probability to remain unscattered after
$2$ steps would be $P^{\rm cl,naive}_{nn}(2)=(1-f)^2$, giving rise to
a {\it quantum} probability $|A^{\rm naive}_{nn}(2)|^2=(1-f)^4$.
This is the same
probability that we would obtain by simply iterating the diagonal part of the
evolution matrix, i.e. by approximating
$(U^2)_{nn} = \sum_{n'}U_{nn'}U_{n'n} \approx U_{nn}U_{nn}$. Of course,
the true probability to remain in state $p$ after two steps is
$p$-dependent: for most values of $p_n$, namely $|p_n|>f$, entirely
different parts of the wavefunction are scattered at each of the two steps,
so the classical probability to remain is
$P^{\rm cl}_{nn}(2)=1-2f<P^{\rm cl,naive}_{nn}(2)$. On the other hand, for
$p$ very close to zero, most of the probability that would be
scattered at the second step has already been lost on the first
scattering event, so
$P^{\rm cl}_{nn}(2) \approx 1-f > P^{\rm cl,naive}_{nn}(2)$.

The analysis can be extended easily to longer times.
The quantum probability to remain after $t$ steps
is still given, to leading order
order in $\hbar$, by 
\begin{equation}
\label{stprob}
|(U^t)_{nn}|^2=|A_{nn}(t)|^2=|P^{\rm cl}_{nn}(t)|^2 \,,
\end{equation}
where naively (in the absence of correlations)
\begin{equation}
\label{pnaive}
P^{\rm cl,naive}_{nn}(t)=(1-f)^t \,.
\end{equation}
The true value of $P^{\rm cl}_{nn}(t)$ for $t>1$ is $p$-dependent; in quantum
mechanics, this $p_n-$dependence can be understood in terms of amplitude
that diffracts from $p_n$ to a nearby channel in one step and diffracts
back into $p_n$ during a following scattering event. The extra amplitude 
coming from diffracting back and forth between nearby channels can 
add in or out of phase with the ``naive" contribution. As we found 
previously, the probability to scatter into a nearby channel after one step
is $f(1-f)$; this is comparable to the probability $f$ for scattering
into a classically distant channel and being completely lost from the system
as far as the short-time return probability is concerned.

Notice that the short-time return probability of
Eq.~\ref{stprob} is completely
independent of the shape of the bump $V(q)$ (as long as the bump amplitude $K$
is chosen to be $O(1)$ so as to allow scattering into many distant channels).
In fact, we can compute the short time return probability in a simplified
model where the non-zero part of the potential is replaced by an absorber, and
pieces of the probability density simply get subtracted from the system.
For given $t$, we then have a distribution of the quantum probabilities
to remain $|A_{nn}(t)|^2$. We easily see that the {\it fastest} possible
decay of the initial state $p_n$ is obtained for $p_n=f$, where an entirely
untouched piece of the wavefunction is absorbed at each step:
\begin{equation}
\label{fastdec}
P^{\rm cl,min}_{nn}(t)=1-tf\,,
\end{equation}
for $t<1/f$. The largest values of $P^{\rm cl}_{nn}$ arise from
$p_n$ near zero, as described above, and also from $p_n$ that
are near simple fractions like $1/2$ or $2/3$. These slowly-decaying momentum
channels give rise to the most non-ergodic long-time quantum behavior, as we
shall see below.

For each channel $p_n$, then, we can compute the quantum short-time
autocorrelation $A_{nn}(t)$; it is given by
the square root of the quantum probability to stay (Eq.~\ref{stprob}),
times the phase
accumulated from the (free) quantum dynamics:
\begin{equation}
\label{afrompcl}
A_{nn}^{\rm short}(t)=
\langle p_n|U^t|p_n\rangle= e^{-ip_n^2t/2} P^{\rm cl}_{nn}\left(|t|\right) \,.
\end{equation}
This holds for
both positive and negative short times (note $A(-t)=A^\ast(t)$ by
unitarity). For a typical momentum $p_n$, $A_{nn}^{\rm short}(t)$
has a decay time
of $O(1/f)$; upon Fourier transforming we obtain a short-time spectral envelope
$S^{\rm short}(E)$ centered at
quasienergy $p_n^2/2$ and with width of order $f$.
Specifically, using the naive estimate of Eq.~\ref{pnaive} and taking the
bump size $f$ to be small, we obtain a
Lorentzian short-time envelope
\begin{equation}
\label{snaive}
S_n^{\rm short,naive}(E)= {2 f \over f^2+\left(E-{p_n^2 \over 2}\right)^2}
\end{equation}
for $\left|E-{p_n^2 \over 2}\right| \ll 1$.

\subsection{Long-time behavior and stationary properties}

\subsubsection{Scaling properties} 

At times long compared with $1/f$, most of the initial amplitude in
a typical channel $p_n$ will have been scattered by the bump, and the
return amplitude $A_{nn}^{\rm long}(t)$ will be given semiclassically by a
sum over many nontrivial paths (the relative phases between the paths being
of course $\hbar-$dependent). As discussed in Section~\ref{stdyn}, and
in completely analogy with nonlinear scarring, these long-time
recurrences are given by independent Gaussian random variables, convoluted
with the short-time dynamics $A_{nn}^{\rm short}(t)$. The full local spectrum
$S_n(E)$ (Eq.~\ref{ldos}) is a line spectrum with individual intensities
$P_{n\xi}=N|\langle n |\xi\rangle|^2$ (where we have 
denoted $\langle p_n |$ as $\langle n |$) being given by a $\chi^2$
distribution,
weighted by the height of the linear envelope at energies $E_\xi$:
\begin{equation}
P_{n\xi}= r_{a\xi} S_n^{\rm short}(E_\xi)\,.
\end{equation}
Here $r_{a\xi}$ are independent $\chi^2$ variables with mean unity (see
Eq.~\ref{linespec}).
The expected local IPR (Eq.~\ref{ipra})
is given by a product of a factor associated with the
short-time envelope and a factor $(F=3)$ associated with the spectral
fluctuations $r_{a\xi}$ under the envelope:
\begin{equation}
\label{iprfromshort}
{\rm IPR}_n = 3 \langle\left(S_n^{\rm short}(E)\right)^2\rangle_E=
3\sum_{t=-\infty}^{+\infty} |A_{nn}^{\rm short}(t)|^2\,.
\end{equation}
Using the naive short-time dynamics of Eq.~\ref{pnaive}, we obtain
an estimate for the typical IPR:
\begin{equation}
\label{naiveipr}
{\rm IPR}_{n}^{\rm naive} =
3 \times {2-2f-f^2 \over 2f-f^2} \approx {3 \over f}\,.
\end{equation}
Using the upper bound we obtained in Eq.~\ref{fastdec} on the rate of
short-time decay of an initial momentum channel,
we also have a lower bound on the local IPR:
\begin{equation}
\label{minipr}
{\rm IPR}_n^{\rm min} = 3 \times {2 \over 3f} = {2 \over f}
\end{equation}
Notice that this lower bound is for moderate $f$ already larger than
the RMT expectation of $3$.

For a given value of $f$, we may use our knowledge of the short-time
classical dynamics and Eqs.~\ref{afrompcl},~\ref{iprfromshort} to obtain
a distribution of the local inverse participation ratios ${\rm IPR}_n$.
For $f \ll 1$, the decay time of the classical autocorrelation
function $P_{nn}^{\rm cl}(t)$ scales with $1/f$, so we expect the IPR
distribution ${\cal P}_f$ to scale likewise:
\begin{equation}
\label{scaleprob}
{\cal P}_f({\rm IPR}_n=x)=f {\cal P} (fx)
\end{equation}
for some function ${\cal P}(y)$. From Eq.~\ref{minipr} we have a lower bound
on possible IPR's for small $f$, i.e.
\begin{equation}
\label{lowerp}
{\cal P}(y) = 0 \; {\rm for} \; y<2 \,.
\end{equation}
Using the naive estimate of Eq.~\ref{naiveipr} for the IPR
at a ``typical" value of the momentum, we determine
that the median of the distribution ${\cal P}(y)$ should be in the 
neighborhood of $3$. This discussion  of the IPR
distribution has been very general; however the details of the function
${\cal P}(y)$ may in fact depend on classical system parameters other than
the bump size $f$. For example, in the equations of motion Eq.~\ref{eqmo}
we could have replaced the free evolution in the second line with
\begin{equation}
q \to \tilde q = q + \alpha \tilde p \; {\rm mod} \; 1 \,,
\end{equation}
making the elapsed time $\alpha$ between kicks an arbitrary parameter.
(In the Sinai billiard system, the parameter $\alpha$ corresponds roughly
to the length-to-width ratio of the rectangular billiard.) The detailed
properties of the IPR distribution ${\cal P}(y)$ will then depend on the
values of classical parameters such as $\alpha$, while results such as
Eq.~\ref{lowerp} apply more generally to the entire class of Sinai-type
systems. 
Below, in Figs.~\ref{kickfig1} and \ref{kickfig2},
we present the actual classically
computed function ${\cal P}(y)$ for the Sinai kicked map with $\alpha=1$;
there ${\cal P}(y)$ is compared with statistics collected for the
corresponding quantum system.

\subsubsection{Tail of the IPR distribution} 

First, we discuss another important qualitative feature of the IPR
distribution, namely the long tail of ${\cal P}(y)$ coming from momentum
channels $p_n$
which are near simple fractions and thus decay on a time scale longer
than the typical $O(1/f)$ steps.
Consider a very small initial momentum $|p_n| \ll f$. As we saw in
the discussion immediately preceding Eq.~\ref{stprob}, 
only a very small fraction
of the remaining amplitude in $|p_n\rangle$ is scattered during each kick 
following the first one, because the part of the wave
which has not yet been scattered
shifts very little in position
space between kicks. Explicitly, the classical probability (and thus the
quantum amplitude) to remain after $t$ steps is given by
\begin{equation}
|A_{nn}^{\rm short}(t)|=P_{nn}^{\rm cl}(t)=1-f-(t-1)p_n
\end{equation}
for $1 \le t \le (1-f+p_n)/p_n$. Thus the decay time for the initial state
$|p\rangle$ scales as $1/p_n$, and the inverse participation ratio
${\rm IPR}_n$ scales likewise (compare Eq.~\ref{iprfromshort}).
These IPR's are large compared to the
$O(1/f)$ IPR's obtained for typical channels (yet small compared to
the $O(N)$ IPR's which characterize bouncing ball states).

Similarly, if we choose a momentum channel which lies near a simple
fraction, $p_n={m \over \ell} +\epsilon$ (for $\ell \epsilon< f<1/\ell$),
then following the first $\ell$ kicks,
a fraction $\ell \epsilon$ is scattered after each successive kick,
and the decay
time (and IPR)
for such a channel therefore scales as $1/\ell \epsilon$. 
We dub these special momentum channels the ``near-bouncing ball"
trajectories.
We can now
easily estimate
the fraction of channels with IPR greater than some number $x$, where
$x \gg 1/f$. All channels within $1/x$ of zero satisfy this condition, as
do those within $1/\ell x$ of a simple fraction $m/\ell$. Now we note that for
a typical integer $\ell$, a finite fraction of integers $m=1 \ldots \ell$ are
relatively prime to $\ell$; thus from each value of $\ell$ we obtain a fraction
$O(\ell \times {1 \over \ell x}) = O(1/x)$ of channels with IPR expected to be 
greater than $x$. Adding up contributions from all values of $\ell$
between $1$ and $1/f$, we have a cumulative probability $O(1/fx)$ for 
${\rm IPR}_n$ to be greater than $x$, or
\begin{equation}
\label{invsquare}
{\cal P}({\rm IPR}_n =x )  \sim {1 \over f x^2} 
\end{equation}
for $x \gg 1/f$.
We see that the parameter $f$ enters in the expected way, and the
tail of the scaling distribution (compare Eq.~\ref{scaleprob}) is then given by
\begin{equation}
\label{invsqscal}
{\cal P}(y) \sim 1/ y^2
\end{equation}
for $y \gg 1$.

\subsubsection{Failure of channel level ergodicity } 

From the inverse square form of the IPR distribution tail in
Eq.~\ref{invsquare}, it would appear that the mean value of the IPR
diverges logarithmically for these systems. However, we notice that
at fixed energy the possible IPR is bounded above by the total number
of channels $N$ (this being the IPR for a pure bouncing ball state), and so
we have
\begin{equation}
\label{logmean}
\langle {\rm IPR}_n\rangle_n = \langle P_{nn} \rangle \sim {\log N \over f} \,.
\end{equation}
We see from Eq.~\ref{logmean} that the mean inverse participation ratio in the
kicked Sinai systems diverges logarithmically with increasing energy
(or decreasing $\hbar$); thus the wavefunctions are becoming less and less
ergodic at the single-channel scale even
as the classical limit is approached,
despite the ergodicity of the corresponding classical mechanics.
The situation
is more surprising here than in the slow ergodic systems~\cite{wqe}, as
in the present case
the Lyapunov exponent is positive and the Sinai billiards have long been
considered a prototypical example of strong classical ergodicity and mixing.
We also note that the logarithmically increasing mean IPR in
Eq.~\ref{logmean} is due not to the bouncing ball states (the fraction of these
scales as $1/N$ and thus their contribution to the mean is $N$-independent),
but rather to the ``near-bouncing ball" channels, whose decay time
is large compared to the typical decay time $1/f$ but still small compared
to the Heisenberg time $N$ at which individual eigenstates are resolved.
Each such channel contributes to many eigenstates of the system, but only
a small fraction of {\it all} the available eigenstates.

Having made predictions about the structure of the IPR distribution
for Sinai-type systems ($2/f$ lower cutoff, $O(1/f)$ median,
$O({\log N \over f})$ mean, inverse square tail), we now proceed
to perform a similar analysis for the other statistical
quantities discussed in Section~\ref{measures}.
As discussed previously, the details (factors of order one) of
the various distributions and statistical averages will be system-dependent,
and can be computed explicitly for any specific Sinai-type system (as
we will do in the following section). What we are interested in here is
the universal scaling behavior of wavefunction
structure with the bump size $f$ and the wavelength $1/N$.

We consider first the individual
wavefunction intensities $P_{n\xi}$ in the momentum
basis. For the typical momentum $p_n$, we have seen that
the smoothed spectrum 
$S_n^{\rm short}(E)$ has a Lorentzian
peak of height scaling as $1/f$ and width scaling
as $f$, centered on the optimal energy $E_n={1 \over 2}p_n^2$ (see
Eq.~\ref{snaive}). Far from $E_{n}$, the smoothed spectrum levels off
to a height of order $f$, leading to the typical behavior
\begin{equation}
P_{n\xi}^{\rm median} \sim f \,.
\end{equation}
Notice that the median intensity is much smaller than the mean
(cf. Eq.~\ref{meanpnxi}).

Even for the most anomalously localized channels, the minimum value
of $S_n^{\rm short}(E)$ never falls below $O(f)$; the smallest 
values of $P_{n\xi}$ must therefore arise from $\chi^2$ fluctuations
multiplying this typical intensity. For complex wavefunction amplitudes
$\langle p|\xi \rangle$, this implies
\begin{equation}
{\cal P}(P_{n\xi}=x) \sim {1 \over  f} \exp(-x/f) \;\; [x \ll f]\,,
\end{equation}
and a corresponding expression is obtained in the real case, where
$P_{n\xi}$ is a $\chi^2$ variable of mean $O(f)$ and {\it one}
degree of freedom:
\begin{equation}
\label{ppne}
{\cal P}(P_{n\xi}=x) \sim {\exp(-x/ 2 f) \over \sqrt{2 \pi x f}} 
\;\;\; [x \ll f] \,.
\end{equation}
The mean of the intensity distribution is of course fixed by
normalization:
\begin{equation}
\label{meanpnxi}
\langle P_{n\xi}\rangle=1 \,,
\end{equation}
where once again the averaging $\langle \cdots \rangle$ can be thought of as an
average over eigenstates $|\xi \rangle$, momentum channels $|p_n \rangle$,
or over some ensemble of Sinai-type systems (where e.g. the shape of the 
bump can be varied while preserving its total size $f$). 

\subsubsection{Tails of intensity and transport measures}
 
Lastly, we turn
to the tail of the intensity distribution, which we expect to result from
large values of the smoothed spectrum $S^{\rm short}(E)$. (Fluctuations of
the full spectrum $S(E)$ around its smoothed value have a $\chi^2$
form. The probability of obtaining an intensity $P_{n\xi}$
large compared to the short-time prediction $S_n^{\rm short}(E_n)$ is
exponentially small.) As previously discussed, a fraction $O(1/fx)$
of all momentum channels $|p\rangle$ have a peak in the spectrum
$S_n^{\rm short}(E)$ of height $ \ge x$, and the width of such a peak is
then $O(1/x)$. Therefore a fraction $O(1/f x^2)$ of all intensities 
$P_{n\xi}$ satisfy the condition $S_n^{\rm short}(E_n) \ge x$, and
since the fluctuations in $P_{n\xi}$ around this smoothed value are of order
unity, we obtain
\begin{equation}
\label{ppntail}
{\cal P}(P_{n\xi} =x) \sim { 1 \over f x^3 }  \;\;\; [x \gg {1 \over f}] \,.
\end{equation}

The eigenstate-basis IPR measure ${\rm IPR}_\xi= P_{\xi\xi}$
(Eq.~\ref{pnndef}),
which measures
the inverse fraction of channels in which a given eigenstate lives may
be studied in a manner very similar to the channel-basis measure
${\rm IPR}_n=P_{nn}$. From Eq.~\ref{logmean}, we already know the
{\it mean} value
of the ${\rm IPR}_\xi$ distribution:
\begin{equation}
\langle{\rm IPR}_\xi\rangle = \langle P_{\xi\xi}\rangle
 \sim {\log N \over f} \,.
\end{equation}
We proceed to study the structure of the ``typical" wavefunction
$|\xi \rangle$.
From Eq.~\ref{ppntail}, we know that given some eigenstate $|\xi \rangle$,
the probability that is has intensity $ \ge x$ at any particular momentum
$|p \rangle$ is $O(1/f x^2)$. If we assume the overlaps of $|\xi\rangle$
with the different momentum states to be uncorrelated, and notice that there
are a total of $N$ momentum channels to overlap with, we see
that for $x \le \sqrt{N/f}$ there will {\it generically} be at least one
momentum $|p_n\rangle$ such that $P_{n\xi} \ge x$.
We now compute the contribution
to $P_{\xi\xi}$ from all intensities
$P_{n\xi}=x$ between $1/f$ and $\sqrt{N/f}$:
\begin{eqnarray}
P_{\xi\xi} & = & \langle P_{n\xi}^2 \rangle =
 \int dx \; x^2 \; {\cal P}(P_{n\xi}=x) \nonumber \\
& \ge & \int_{1/f}^{\sqrt{N/f}} dx \; x^2 {1 \over f x^3} \sim 
{\log N  \over 2f}
\end{eqnarray}
(recalling $N \gg 1/f$). Thus we see that not only the mean, but also the
inverse participation ratio for the {\it typical} wavefunction tends to
infinity in the classical limit:
\begin{equation}
{\rm IPR}_\xi^{\rm median} \sim {\log N  \over 2f} \,.
\label{logtyp}
\end{equation}
The tail of the ${\rm IPR}_n$ distribution arises from the rare intensities
$P_{n\xi} \gg \sqrt{N/f}$, and using Eq.~\ref{ppntail} is easily seen to
take the form
\begin{equation}
{\cal P}({\rm IPR}_\xi=x) \sim { 1\over f x^2} \,.
\end{equation}

Having analyzed the statistical structure of individual wavefunctions
in Sinai-type systems, we can now proceed to examine the quantum transport
behavior. Specifically, we focus on the long-time transport probability
$P_{nn'}$ between two channels $p_n$ and $p_{n'}$, as introduced previously
in Eqs.~\ref{pabdef},~\ref{pabdyn}. For two {\it typical} channels
$p_n$ and $p_{n'}$, each of the two smoothed local densities of states has
the form of a peak of height $1/f$ and width $f$ centered around some energy,
and then falls off to a value of $O(f)$ far from that energy
(see Eq.~\ref{snaive}). Since the two peaks are generically centered
at different energies, $|E_n-E_{n'}| \gg f$,
we easily see that the overlap between the two
envelopes is $O(f)$:
\begin{equation}
P_{nn'}^{\rm median} \sim f \,.
\end{equation}
Of course we also know the mean value of this distribution by construction:
\begin{equation}
\langle P_{nn'}\rangle =1 \,.
\end{equation}
Large values of the transport measure $P_{nn'}$ arise from those $p_n$
and $p_{n'}$ for which the two spectral envelopes $S^{\rm short}_n(E)$
and $S^{\rm short}_{n'}(E)$ are both anomalously tall and narrow, and which
also have significant overlap with each other. Explicitly, in order to obtain
a value $P_{nn'} \ge x$ for large $x$, we require ${\rm IPR}_n \ge x$,
${\rm IPR}_{n'} \ge x$, and also $|E_n-E_{n'}|<{1 \over x}$. The combined 
probability for these three unlikely events scales as
\begin{equation}
{\cal P}(P_{nn'} \ge x) \sim { 1\over f x^2} \times {1 \over f x^2} \times
{1 \over x} \,,
\end{equation}
so
\begin{equation}
{\cal P}(P_{nn'} = x) \sim { 1 \over f^2 x^4}\;\;\; [x \gg {1 \over f}] \,.
\label{quar}
\end{equation}
This a very quickly decaying tail compared to the one obtained previously
for the inverse participation ratio $P_{nn}$ (compare Eq.~\ref{invsquare});
thus transport efficiency for this class of systems is much less anomalous
than the structure of individual wavefunctions. This makes sense intuitively
and is also consistent with the findings for slow ergodic systems in
Ref.~\cite{wqe}.

Finally, the remaining measure we must consider is the final-state-averaged
transport efficiency $Q_n$ for initial state $p_n$ (introduced in
Eq.~\ref{qdef}; see also Eqs.~\ref{pabdef},~\ref{pabdyn} for the definition
of $P_{nn'}$, 
the long-time probability of getting to channel
$p_{n'}$ from channel $p_n$).
The quantity $\langle Q_n \rangle$, as well as the typical value
of $Q_n$, will be dominated by the Lorentzian
envelopes governing typical intensities $P_{n\xi}$:
\begin{eqnarray}
\label{lorentz} 
 P_{nn'} &=& {1 \over N} \sum_\xi P_{n\xi}P_{n'\xi} \nonumber \\
         &\sim& 
      \int \ {dE\over 2\pi}
       \left({2f\over f^2 + (E-p_n^2/2)^2}\right)
                  \left({2f\over f^2 + (E-p_{n'}^2/2)^2}\right) \nonumber \\
         & = & \left({4f \over 4f^2 + (p_n^2/2 - p_{n'}^2/2)^2}\right)\,,
\end{eqnarray} 
where in the second line we have inserted the typical intensity in
channel $p_n$
of a state $|\xi\rangle$ with energy $E$ (from Eq.~\ref{snaive}).
Now 
\begin{eqnarray}
Q_n &=&{1 \over N}\sum_{n'} P_{nn'}^2 \nonumber \\
& \sim &\int  \ {d(p_{n'}^2/2) \over 2 \pi}
\left ({4f\over 4f^2 + (p_n^2/2-p_{n'}^2/2)^2}\right )^2 = {1 \over 2f}\,.
\end{eqnarray} 
The mean and the median both scale as
\begin{equation}
\langle Q_n \rangle \sim Q_n^{\rm median} \sim {1 \over f};
\label{qpred}
\end{equation}
and furthermore in the classical limit $N \to \infty$ it is exceedingly
difficult to obtain values of $Q_n$ either small or large compared to
$O(1/f)$.
For {\it almost any} initial channel $p_n$, the
fraction of final channels $p_n'$ to which one can be transported at long times
is $O(f) \ll 1$. Bouncing-ball (free propagation)
channels of course have even less coupling
to other momentum states (roughly speaking, they couple to themselves only,
$P_{nn} = O(N)$ and thus $Q_n = O(N)$), but these constitute
a vanishing fraction of all channels in the classical limit.

\section{Numerical Tests in Sinai Kicked Maps}
\label{numtests}

\begin{figure}
\centerline{\epsfig{figure=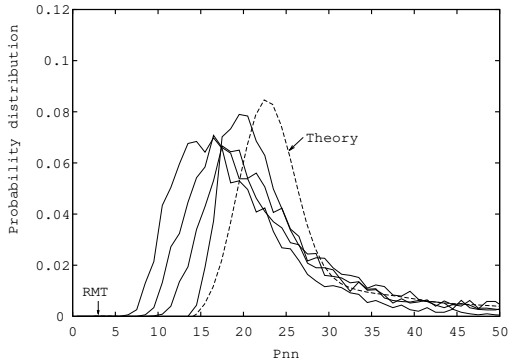,width=7cm}}
\vskip 0.05in
\caption{The distribution of the inverse participation ratio ${\rm IPR}_n=
P_{nn}$ is plotted for bump size $f=0.1$. From left
to right, the four solid curves represent data for $N=100,$ $200$, $400,$
and $1600$. The theoretical prediction (see text) is represented by dashes,
while the random matrix prediction is ${\rm IPR}=3$.}
\label{kickfig1}
\end{figure}

We proceed to a  numerical study of the structure of wavefunctions in the
Sinai kicked systems, focusing on those statistical properties which we have
treated theoretically in the preceding section. We begin by considering the
distribution of inverse participation ratios ${\rm IPR}_n$ (Eq.~\ref{ipra}),
each of which measures
the inverse fraction of eigenstates having significant intensity at some 
momentum channel $p_n$.
The bump size $f$ is fixed at the moderate value
of $0.1$, which is small compared to the system size of unity,
yet large compared to
wavelengths $1/N \le 0.01$ which we are going to consider.
In Fig.~\ref{kickfig1}, the IPR distribution
${\cal P}_f({\rm IPR}_n)$ is plotted (solid curves)
for several values of the quantum
wavelength:
$N=100,$ $200,$ $400,$ and $1600$. In each case, an ensemble has been
constructed by varying the bump height parameter $K$ in Eq.~\ref{kickpot};
each realization with $K=O(1)$ is expected to have the same wavefunction
statistical properties, as discussed in the preceding section.

Also plotted
as a dashed curve 
in Fig.~\ref{kickfig1} is a classical prediction for ${\cal P}_f({\rm IPR}_n)$.
This quantity is obtained by taking a random sample of initial
momenta $p_n$, and for each of them computing classically the probability 
$P_{nn}^{\rm cl}(t)$ to remain unscattered after $t$ bounces. Given 
the short-time classical behavior $P_{nn}^{\rm cl}(t)$, we use
Eqs.~\ref{afrompcl},~\ref{iprfromshort} to predict the expected quantum
IPR for that momentum channel $p_n$, eventually leading to the distribution
shown by the dashed curve. Of course, this is a semiclassical ($N \to \infty$)
prediction; in particular, it ignores fluctuations in the IPR which
result from summing over a finite number of eigenstates in Eq.~\ref{ipra}
(even in RMT, fluctuations around the mean value of $3$ are
expected for finite $N$).

Indeed, we see in Fig.~\ref{kickfig1} that the quantum IPR distribution does
appear to approach the classically predicted distribution as $N$
gets large; the convergence
with $N$ will be studied more quantitatively
below in Fig.~\ref{iprnf}. By the time we reach $N=1600$, the peak of the
distribution is within $10 \%$ of the classically expected value, and is
shifted by a factor of seven from the naive random matrix prediction.
We also notice that {\it all} the IPR's in our
sample are larger than the value of $3$ predicted by random matrix theory,
and most are larger by a factor of at least five:
this is unmistakable evidence of strong deviations from microscopic quantum
ergodicity in the kicked Sinai systems.

\begin{figure}
\centerline{\epsfig{figure=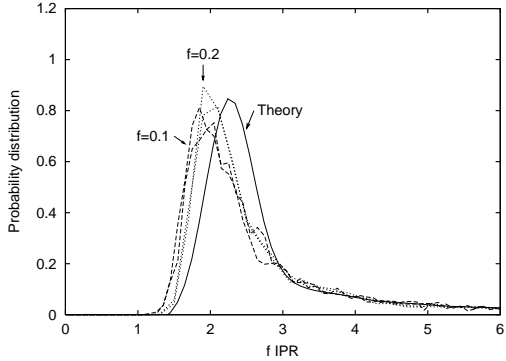,width=7cm}}
\vskip 0.05in
\caption{The distribution of the scaled inverse participation ratio 
$f \cdot {\rm IPR}$ is plotted for $f=0.1$ (dashed curves) and
$f=0.2$ (dotted curves). $N=1000$, and for each value of $f$ two
distributions are plotted corresponding to different bump shapes
(see text).}
\label{kickfig2}
\end{figure}

Next, in Fig.~\ref{kickfig2} we fix the total number of channels
at $N=1000$, and study the IPR distribution for various values of the bump size 
$f$. Guided by the predicted scaling relation of Eq.~\ref{scaleprob}, we choose
to plot the distribution of the scaled quantity
$f \cdot {\rm IPR}_n$ for each value of bump size $f$. For each of
$f=0.1$ (dashed curves) and $f=0.2$ (dotted curves),
two distributions are plotted: one for
the original kick potential of Eq.~\ref{kickpot}, and the other for a modified
kick potential
\begin{eqnarray}
\label{kickpot2}
V(q)&=&-{K \over 2f} \left[\left(\left |q-{1 \over 2}\right |+
{f \over 2}\right)^2
 - f^2\right] \;
 \; {\rm for} \; \left|q-{1 \over 2}\right| < {f \over 2} \nonumber \\
&=& 0 \;\; {\rm otherwise.}
\end{eqnarray}
The latter potential has a kink at $q=1/2$, causing a discontinutiy in the 
classical dynamics. We see from Fig.~\ref{kickfig2} that the choice of kick
potential (Eq.~\ref{kickpot} or Eq.~\ref{kickpot2}) has no significant effect
on the IPR distribution, as long as the bump size $f$ is fixed,
confirming the universality predicted in the previous section. In particular,
we notice that the flat part of the potential [$V'(q=1/2)=0$] in
Eq.~\ref{kickpot}, which scatters any incoming channel back into itself,
has no discernible effect on quantum localization at the energies
under consideration.

The classical prediction for the scaling distribution
${\cal P}(f \cdot {\rm IPR})$ is also plotted in Fig.~\ref{kickfig2}
for comparison (see solid curve).
We see very good agreement among the $4$ sets of quantum data
at $f=0.1$ and $f=0.2$; similar scaling behavior with $f$ is observed
for the billiard system in Fig.~\ref{calP}.
Again, the
slight discrepancy (around $10 \%$) between the numerical
data and the classical
prediction may be attributed to the finiteness of the energy. At
these energies, the minimum observed value of the IPR appears
to be near $1.5/f$, in contrast to the $2/f$ semiclassical limit
prediction of Eq.~\ref{minipr}.

\begin{figure}
\centerline{\epsfig{figure=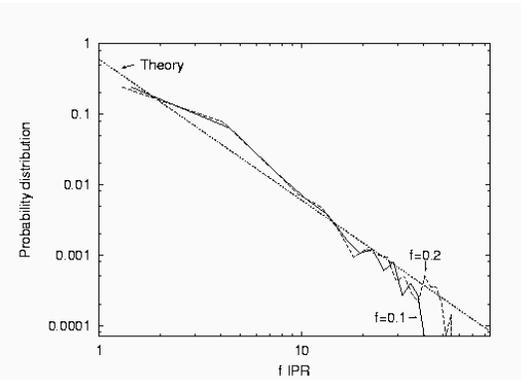,width=7cm}}
\vskip 0.05in
\caption{The tail of the IPR distribution is plotted for $N=1000$, with
$f=0.1$ (solid curve) and $f=0.2$ (dashed curve). The theoretically
predicted $0.6/x^2$ behavior (Eq.~\ref{invsqscal}) appears a dotted
line.}
\label{kickfig2b}
\end{figure}

In the tail, we predict (Eq.~\ref{invsquare}) the
inverse square behavior ${\cal P}(x) \approx c/fx^2$ for
the IPR distribution, where the
constant $c$ can be determined  to be $0.6$ through a detailed
classical analysis of this system as described above. The $c/y^2$
tail for $y=f\cdot {\rm IPR}$
is indeed observed in Fig.~\ref{kickfig2b}, where the
prediction appears as a dotted line on the log-log plot, while the
solid and dashed curves represent $f=0.1$ and $f=0.2$, respectively.
This data was again taken for $N=1000$, and we see the
power-law behavior persist to IPR's of about $300$, where the IPR
becomes comparable to the total number of channels and the theory
naturally breaks down.

We recall that this breakdown of the inverse-square law at ${\rm IPR} \sim
N$ leads to the prediction of mean IPR growing logarithmically with
$N$: $\langle {\rm IPR} \rangle \approx 0.6 \log N +{\rm const}$ (see
Eq.~\ref{logmean}). This behavior is indeed observed for $N$ ranging from
$100$ to $1000$; we omit the figure here because an analogous plot
for the billiard system appears in Fig.~\ref{iprnf} in the following
section. The {\it median} $P_{nn}$ shows no such increase with $N$;
it saturates at $\approx 2.35 /f$ independent of $N$ (see Eq.~\ref{naiveipr}).
The median IPR for an eigenstate ($P_{\xi\xi}$), on the other hand, does grow
logarithmically with $N$, but only half as fast
as the mean, in agreement with Eq.~\ref{logtyp}.

\begin{figure}
\centerline{\epsfig{figure=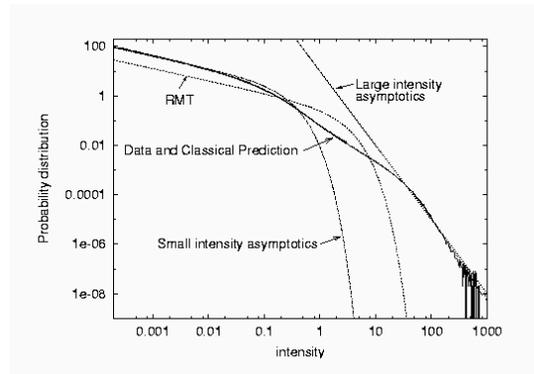,width=7cm}}
\vskip 0.05in
\caption{The distribution of intensities $P_{n\xi}$ for $f=0.1$ and
$N=1000$ (solid curve). The classical prediction (see text) follows
the data very closely (dashed curve). Asymptotics form for the head
(Eq.~\ref{ppne}) and tail (Eq.~\ref{ppntail}) of the distribution
are both drawn using small dashes. For comparison, the Porter-Thomas
distribution of RMT appears as a dotted curve.}
\label{kickfig6}
\end{figure}

We next turn to the distribution of individual wavefunction
intensities. In Fig.~\ref{kickfig6} the distribution of
intensities $P_{n\xi}$ is plotted for $f=0.1$ and $N=1000$ (solid curve).
The classical prediction (obtained as described in the discussion
of Fig.~\ref{kickfig1} above) is plotted as a dashed curve; the
difference between data and prediction is barely visible except in
the very tail where the statistical uncertainty in the data becomes important.
The two analytic asymptotic expressions: $\exp(-x/2f)/\sqrt{2\pi xf}$
for small intensities $x$ (Eq.~\ref{ppne}) and $1/fx^3$ for large $x$
(Eq.~\ref{ppntail}) are also shown in Fig.~\ref{kickfig6}. These
two expressions are valid for $x \ll f$ and $x \gg 1/f$, respectively.
By contrast,
the RMT Porter-Thomas prediction (dotted curve) does not agree
with the data in the head, body, or tail of the distribution.
See also Figs.~\ref{ppn} and \ref{ppnt}, which focus separately on 
the head and tail of the intensity distribution for the billiard
system, and again find good agreement with theory and disagreement
with RMT.

The distribution of transport measures $P_{nn'}$
has also been studied and observed to possess a $1/f^2x^4$ behavior
for $x \gg 1/f$, as predicted in Eq.~\ref{quar}. This data is omitted here 
as very similar behavior is obtained for the billiard in Fig.~\ref{nnnt}.
The overall transport efficiency $Q$ has also been studied
and follows the predicted scaling $Q \sim 1/f$ of Eq.~\ref{qpred}, so
that only a fraction $O(f)$ of all channels are quantum
mechanically accessible at long
times starting in any one initial channel.

\section{Localization in Sinai Billiards}
\label{billiard}

\begin{figure}
\centerline{\epsfig{figure=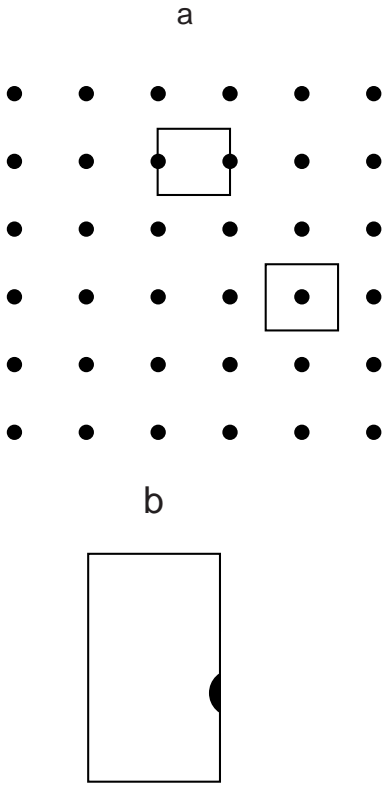,width=5cm}}
\vskip 0.05in
\caption{ a) The Lorenz gas and two choices for a fundamental domain.
b) A Sinai like billiard related to the  Lorenz gas  
}
\label{sinai}
\end{figure}

The Sinai billiard was the first nontrivial dynamical 
system shown to  be ergodic with positive Lyapunov 
exponent\cite{sinai}.  In this sense it is {\it the} paradigm
of chaos.  It is also a unit cell of the  Lorenz gas,
a periodic array of hard disk scatterers (see Fig.~\ref{sinai}a).  

For numerical reasons we investigate a modified Sinai system
with the circular disk off center and jutting only part way into the  billiard;
this is sill a chaotic system 
(see Fig.~\ref{sinai}b).

\subsection{Scattering method} 

\begin{figure}
\centerline{\epsfig{figure=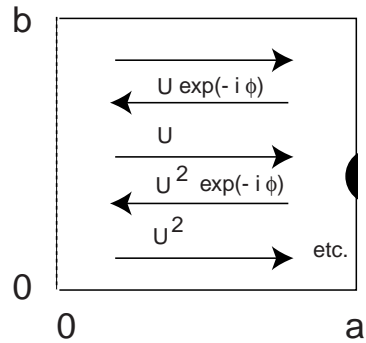,width=5cm}}
\vskip 0.05in
\caption{ The modified Sinai system, with  a
partial disk occupying a variable fraction 
$f$ of the right hand vertical wall.
}
\label{modsinai}
\end{figure}

A scattering system closely connected with both the Lorenz gas and the 
Sinai billiard puts the Sinai 
disk at the end of a corridor of length $a$ (Fig.~\ref{modsinai}). The 
scattering wavefunction can then be expanded as 
\begin{eqnarray} 
\label{scatt}
\Psi(x,y) &=& {1\over \sqrt{k_{n}}}e^{-i k_{n} x}\sin(n \pi y/b)  
\nonumber \\ &-&
\sum_{n'}{1\over \sqrt{k_{n'}}}S_{nn'} e^{-2 i k_{n'}a}
e^{i k_{n'} x}\sin(n' \pi y/b)
\end{eqnarray} 
where for later convenience  we have factored out a phase $\exp(2 i k_{n'} a)$
from the $n'$-th column of the S-matrix.  (If there is no scatterer on the
right hand wall, this makes $S$ the diagonal unit matrix, assuming Dirichlet
conditions there). 
Now suppose that we reflect the scattered wave
from the left wall back towards the right hand side,
in accordance with the  closed  
billiard problem we wish to solve.
This can be done by
imposing a boundary condition at the left wall, which 
need not necessarily be Dirichlet. (We indicate this 
by using  a dashed line to represent this wall in
Fig.~\ref{modsinai}.) 
If the wave is reflected from the left wall at $x=0$, it
  returns
with a new phase $\exp(i\phi)$ given by the boundary
condition at the left wall.
 We   define 
\begin{equation} 
U_{nn'} = S_{nn'}\exp(-2 i k_{n'} a + i\phi) \,.
\end{equation} 
Setting $\psi_n = \exp(-i k_{n} x)\sin(n \pi y/b)/\sqrt{k_n}$,
the net incoming (right-moving) wave is then 
\begin{equation} 
(1 + U + U^2 +\cdots)\psi_n = {1\over 1-U} \psi_n
\end{equation} 
(see Fig.~{\ref{modsinai}}).

Evidently, a bound state can be built up in the billiard if 
 U has an eigenvalue +1.
We can diagonalize the $U$-matrix and consider the
properties of its
eigenstates.   Since $U$ is a unitary matrix, its 
eigenvalues lie on the unit circle.
As we change the phase shift $\phi$ at the 
left wall, the eigenvalues will correspondingly 
rotate  around the unit circle;
each of the N eigenvalues of $U$  (assuming there
are  N open channels) will pass through $+1$ for some $\phi$, 
so that {\it every eigenstate of $U$ is an 
eigenstate of the closed billiard with some boundary condition at the 
left wall and Dirichlet boundary conditions elsewhere}. 

If one is  willing to search through ranges of energies or
of box lengths $a$ one can find a set of eigenstates satisfying 
a particular boundary condition; this is a way of finding eigenvalues and
eigenstates of the billiard with Dirichlet boundary conditions; they are
given by eigenstates of $U$ with eigenvalue $1$\cite{smil,bogo}.
 However here we do not seek the
  Dirichlet solutions, since they are not  special as far as their
localization
in the channel space (this has been tested numerically).  This is 
of  great value in gathering the statistics needed here.

\begin{figure}
\vskip .2in
\centerline{\epsfig{figure=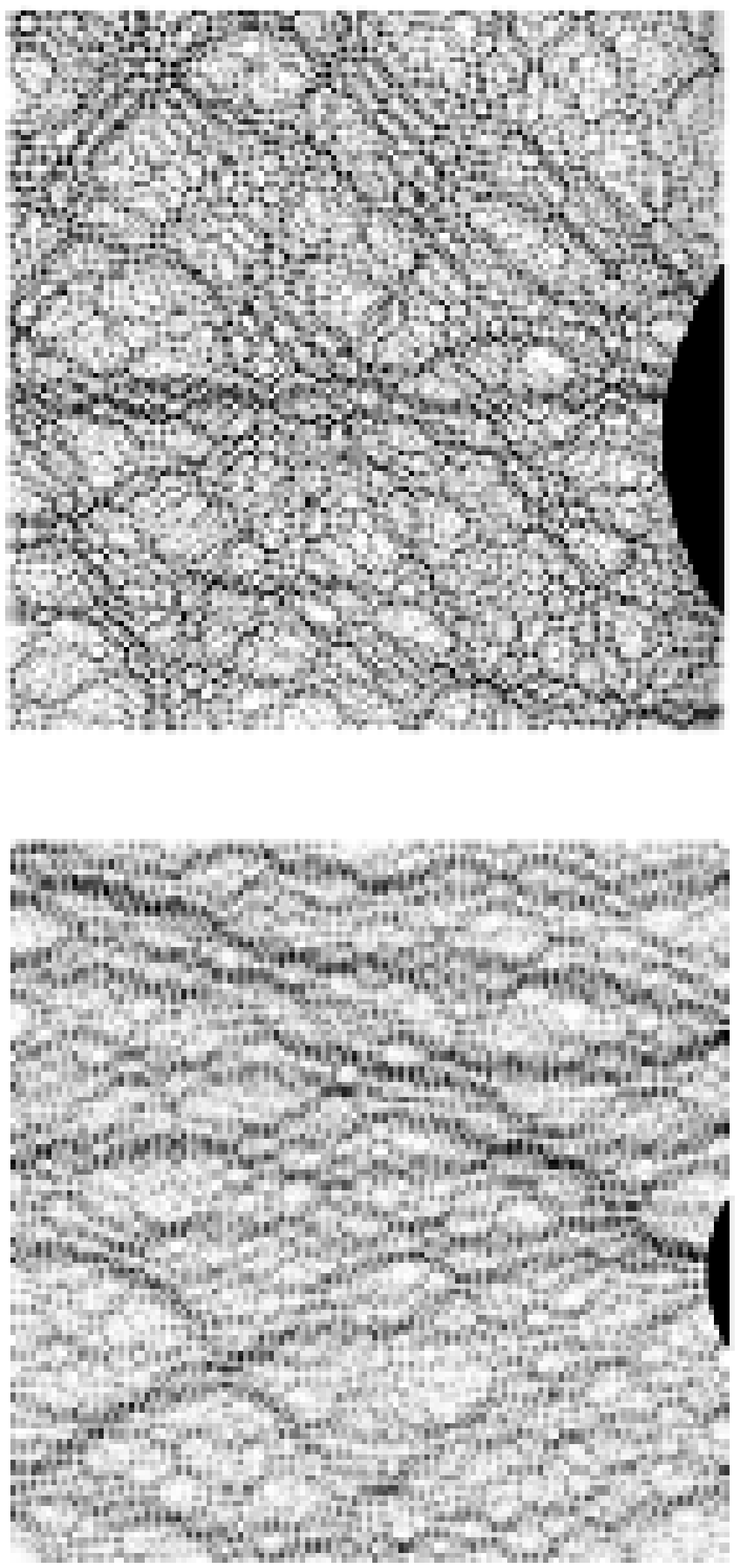,width=10cm}}
\vskip 0.05in
\caption{Two typical eigenstates of the $S$-matrix for
the Sinai-like scattering system.}
\label{states}
\end{figure}

Two typical
 eigenstates of the $U$-matrix are
shown in Fig.~\ref{states}; these show fairly obvious non-statistical mixing
of different directions of propagation in the billiard (nonmixing of
channels in the
scattering approach).

\begin{figure}
\vskip .2in
\centerline{\epsfig{figure=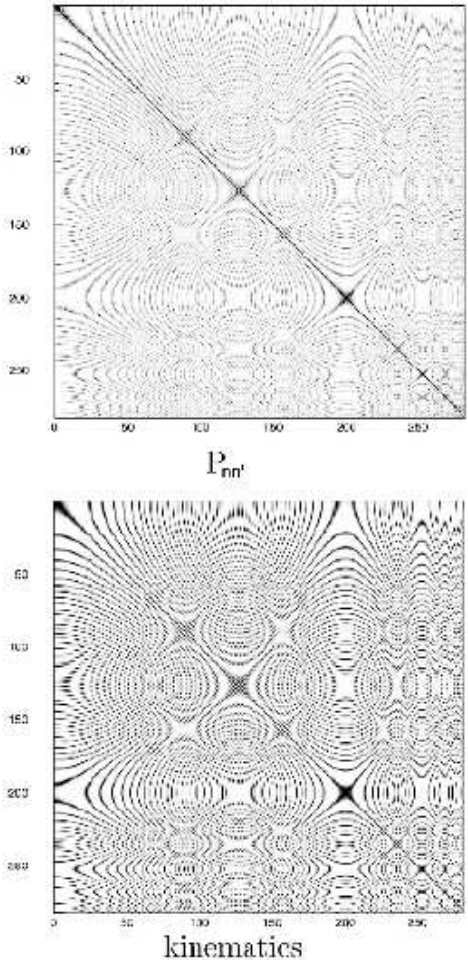,width=7cm}}
\vskip 0.05in
\caption{Top:  long-time transport probability
  between channels $n$ and $n'$ for the $a=b=2 \pi$ (square) billiard, f=0.1
($m=\hbar=1$). Bottom: the fringe pattern from Eq.~\ref{inter}. }
\label{fringe}
\end{figure}

\subsubsection{Patterns in the channel (momentum) transport}

A new twist arises in the channel transport measure $P_{nn'}$,
which we now present.
Heretofore we have been making the point that Gaussian random wavefunction
statistics are 
much stronger than required for SZCdV ergodicity,
and that much coarser randomness
can still lead to ergodic transport classically.
Below, we see that transport in momentum space may 
even be highly organized, but in a way that still
permits coarse grained SZCdV ergodicity.

The density plot of the transport measure $P_{nn'}$ for  a typical
case ($f=0.1$ with 280 open channels, side lengths equal) appears at
the top of Fig.~\ref{fringe}.
A pronounced fringe pattern is evident.  This pattern
changes with the length of the billiard, and as we now show
represents alternating
constructive and destructive interference
due to the phase   factors
 $\exp[2 ik_n a]$ where $a$ is the box length and
$ k_n $ is the horizontal wavevector.  The $S$-matrix itself
 shows none of this fringing, but it is strongly
evident already for $S^2$:
 We have 
\begin{equation} 
 S^2_{nn'} = \sum\limits_{n''} S_{nn''}S_{n''n'},
\end{equation} 
 and since $S$ is diagonally dominated,
 the major contribution to $S^2_{nn'}$ for $n\ne n'$
 is 
\begin{equation} 
\label{inter}
S_{nn}S_{nn'} + S_{nn'}S_{n'n'} =
 S_{nn'}(S_{nn} + S_{n'n'})
\end{equation} 
Of course  $S_{nn}$ and   $S_{n'n'}$ can interfere;  
 these diagonal elements have factors
 $\exp[2 i k_n a]$ and $\exp[2 i k_{n'} a]$, respectively.
 Subsequent iterations reinforce this interference and give
 very sharp preferred channels that one can end up in
 when starting from a given initial channel.
 A plot of 
$$W_{nn'}\equiv \left \vert\exp[2 i k_n a] 
   +\exp[2 i k_{n'} a]\right \vert^{28} $$
appears at the bottom of Fig.~\ref{fringe}, and is seen
to bear a close resemblance to the fringe pattern in $P_{nn'}$
(the exponent $28$ is of course arbitrary and only serves to set
the contrast ratio of the plot). It should
be kept in mind that the fine detail (pixel by pixel) 
of the  intensity modulations present in $P_{nn'}$
are absent in the lower plot, but the overall modulation of the 
regions of large and small $P_{nn'}$ are almost identical.

Interestingly, the special channels which correspond to 
classical free motion (never hitting the obstruction)
show up on the diagonal as hyperbolic
points of  high density.  This may be shown by 
expanding in the channel index
(at least in the lower $n$ region where the Taylor series holds for 
$\Delta n \sim 1$), e.g.
\begin{eqnarray} 
& &\left |\exp[2 i k_{n} a]    +\exp[2 i k_{(n+\Delta n)} a]\right |
 \nonumber \\ &\approx&
\left |1+  \exp[2 i a\ (\partial k_{n}/\partial
n )\ \Delta n ]\right |\,,
\label{interf}
\end{eqnarray} 
where $a$ is the length of the rectangular box.
Since $k_{n} = k_{nx}=\sqrt{2(E-{n^2\pi^2\over 2 b^2})}$ and
$k_{ny}=n\pi/b$, where $b$ is the
height of the box, we have
\begin{equation} 
{\partial k_{n}\over \partial n} = {k_{ny}\over k_{n}} {\pi\over b}\,.
\end{equation} 
Then the interference in Eq.~\ref{interf} is maximally constructive for
\begin{equation} 
{k_{ny}\over k_{n}}   = {m \over  \Delta n}{b\over a}\,,
\end{equation} 
i.e. exactly for the free motion trajectories.
The  special channels correspond with the hyperbolic regions along the diagonal
in  Fig.~\ref{fringe}. 
The near-bouncing ball channels near the free propagation
channels preferentially diffract
symmetrically about these special channels, as evidenced by
the local hyperbolic structure. This is again a consequence
of the interference structure  in Eq.~\ref{inter}.
Essentially, there is a preference to scatter by a multiple
 of  a reciprocal ``lattice'' vector, $(2 a \Delta k  = 2 m \pi)$,
 reminiscent of Bragg scattering from a periodic structure
with lattice constant $a$.  

The dramatic interference pattern is  another interesting
quantum signature  of a short time effect, already evident after
one iteration as explained above.  
It illuminates another variation on the theme of this paper: on scales 
finer than SZCdV, non-Gaussian statistics may prevail.  Here, we 
see a very structured and nonrandom fringe pattern, which however 
varies on a scale proportional to  $\hbar$, doing no harm to the 
Schnirelman limit.
 
\subsection{Numerical method}

The simple method which we use to find the $S$-matrix
makes use of the expansion of Eq.~\ref{scatt}, including
up to 70 or 100 closed channels along with  all the 
open channels (50-500 here) as a basis.  Linear equations
are set up by requiring that $\Psi(x,y)$  vanish 
at up to 1000 points $(x,y)$ along the right hand wall. The basis 
functions already vanish along $y=0$ and $y=b$, which 
is a mixed blessing, since this is also true beyond  the
right hand wall, where this vanishing is unphysical.
Without the inclusion of closed channels the method
does not converge.  The rectangular linear problem 
( $M$ by $N'$, where $N'$ is the total number of channels,
including the evanescent modes,
and M is the number of points along the wall set to 
zero) is then solved by singular value decomposition.
The rationale for inclusion of closed channels is that 
they handle  details of the boundary conditions at the disk
on a scale smaller than a wavelength.
The closed channels are naturally all taken to have total energy equal to
the scattering energy, using $k^2 = k_x^2 + k_y^2$ with 
$k_x$ pure imaginary and $k_x^2 < 0$.  The values of $k_y$ used
were given by the quantized values in the corridor, however
this would  not be necessary if we included additional points along the 
top and bottom walls near the right hand end, and explicitly forced
the total wavefunction to vanish there.

We find that with the restricted basis described above the 
convergence is poor if the disk protrudes too far into 
the billiard.  By keeping the center of the disk well to the right of the 
wall, we are able to get stable results for energies
such that $k d \le 10 \pi$, where $d$ is the distance
the disk protrudes. This means that the obstruction can be made
at least several wavelengths wide in both dimensions, a requirement
that we must satisfy in order to be in the high energy regime.
Typically  the states we study  are in the range
of the 10,000$^{th}$ to 100,000$^{th}$ eigenstate 
of a fixed boundary condition billiard; it is possible 
to go beyond the one millionth state for small 
disks. The range of stability of the method
may perhaps be greatly extended by generalizing
the basis to more flexible evanescent modes, as discussed
above. 

The disk covers a fraction $f$ of the right hand wall.  We take that 
fraction to be between $0.04$ and $0.28$.  In analogy with the map discussed
above, a fraction $1-f$ of the incoming wave
is not scatterered on the first bounce,
approximately independent of the incoming channel. 
The discussion of Section~\ref{short}
holds without modification, including the quantum 
factor of two in the effective cross section of the disk, corresponding
to diffraction into nearby channels.  

The localization of the wavefunction which we now have come  to expect
in channel space ultimately arises from the fact that only a small
fraction of the incoming channel is scattered after each iteration
of the $S-$matrix for small $f$.
The typical scattering channel
presents fresh amplitude to the disk after each bounce, scattering another 
fraction $f$ of the remaining amplitude.  The resulting slow decay
out of the initial channel is already enough to cause gross anomalies 
in the wavefunction statistics, as compared to RMT.
Specifically, this arises from the short-time
induced Lorentzian envelope in the quasienergy spectrum,
as discussed in Sec.~\ref{lorenz}.

Classically there 
are also now a finite number of    angles $\theta$,
with $(a/b)\tan \theta = n/m$ for integer
$m$ and $n$, which never hit the disk.   For channels corresponding
to propagation near these  angles there is a reduction
in scattering out of the initial channel.  These channels are not
true bouncing ball modes, but   near enough to have a strong effect 
on lifetimes.  (``Time'' is 
now the number of iterations of the $U$-matrix.) Again in 
complete correspondence with the discussion above, the tails of various
distributions are governed by these near-bouncing ball
orbits.  

\subsection{Numerical findings--Sinai billiard}

\begin{figure}
\vskip .2in
\centerline{\epsfig{figure=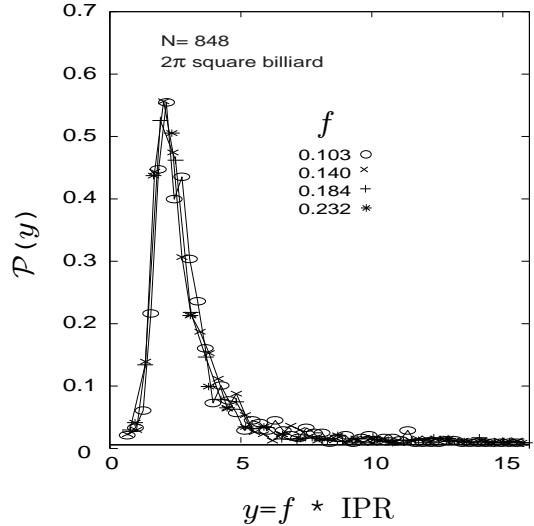,width=7cm}}
\vskip 0.05in
\caption{The probability distribution for the 
IPR's is plotted for various values of $f$, showing the 
predicted scaling behavior. 
}
\label{calP}
\end{figure}

We consider first the 
return probability (inverse participation ratio) measures. 
The scaling relation  ${\cal P}_f({\rm IPR}_n=x)=f {\cal P} (fx)$ was
predicted in Eq.~\ref{scaleprob}; a plot of $f {\cal P} (f x)$
{\it vs} $f x$ for various values of
the disk size $f$ is shown in Fig.~\ref{calP}, confirming this
scaling over the whole  domain of IPR values. We see also from the plot
that the typical IPR in the Sinai system is $\approx 2/f$, which
for the values of $f$ considered is much larger
than the RMT-predicted value of $3$. We also see the expected broad
distribution of IPR's, with $N-$independent width, in contrast
to the RMT prediction that the spread in the IPR distribution should
go to zero as $1/\sqrt N$.

\begin{figure}
\vskip .2in
\centerline{\epsfig{figure=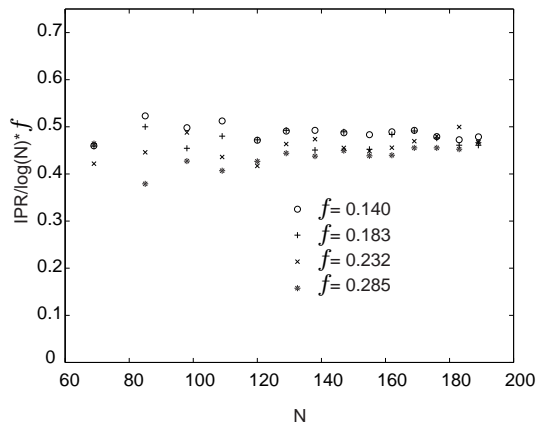,width=7cm}}
\vskip 0.05in
\caption{The average
IPR is plotted, showing the
predicted dependence on N and $f$. 
}
\label{iprnf}
\end{figure}

The tail of the IPR distribution is predicted
(Eq.~\ref{invsquare}) to have the power-law behavior
${\cal P}({\rm IPR}_n =x )  \sim {1 /f x^2} $.
This inverse-square behavior was indeed observed, and
is similar to the same falloff already seen in
Fig.~\ref{kickfig2b} for the kicked Sinai maps.
The power law tail together with the cutoff in the maximum IPR
lead to the prediction of
Eq.~\ref{logmean}, namely 
$\langle {\rm IPR}_n\rangle_n = \langle P_{nn} \rangle \sim {\log N / f} $.
A plot of the the dependence of the average IPR on
N and $f$ is given  in Fig.~\ref{iprnf} , where the 
agreement with Eq.~\ref{logmean} is seen to be excellent.
As predicted, the mean IPR diverges logarithmically away
from its ergodic value of $3$ in the classical limit.

The distribution of {\it small} intensities $P_{n\xi}$ should be  given for 
our S-matrix by Eq.~\ref{ppne}.
\begin{figure}
\vskip .2in
\centerline{\epsfig{figure=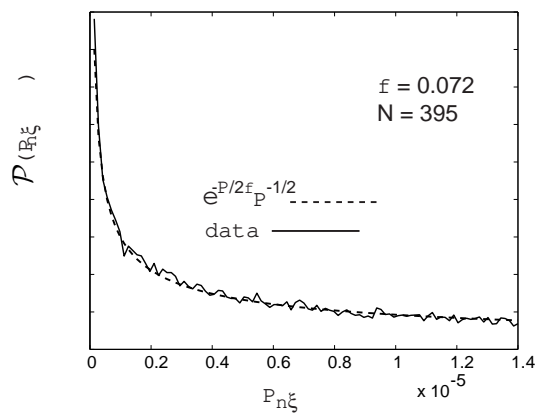,width=7cm}}
\vskip 0.05in
\caption{The distribution of $P_{n\xi}$ is plotted for small
values of  $P_{n\xi}$ and compared with theory,  Eq.~\ref{ppne}.
In this case the bump size is $f=0.072$ and the number of channels
is $N=395$.
}
\label{ppn}
\end{figure}
This  behavior at the low end of the $P_{n\xi}$ distribution is
in very good agreement with the theory (Fig.~\ref{ppn}).

\begin{figure}
\vskip .2in
\centerline{\epsfig{figure=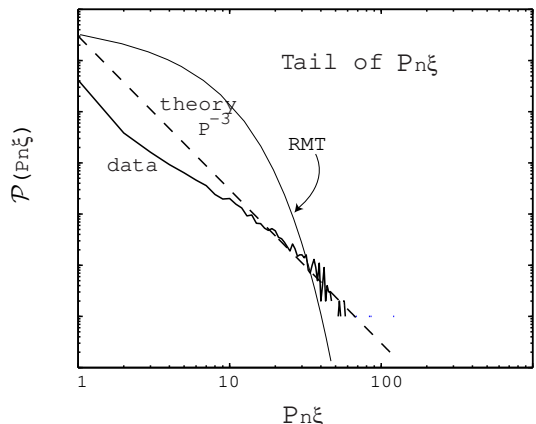,width=7cm}}
\vskip 0.05in
\caption{The tail region of the $P_{n\xi}$ 
distribution  shows
good agreement with the predicted cubic power law, 
for a $2\pi \times 2\pi$
billiard, $f = 0.23$, 226 channels.  
}
\label{ppnt}
\end{figure}

\begin{figure}
\vskip .2in
\centerline{\epsfig{figure=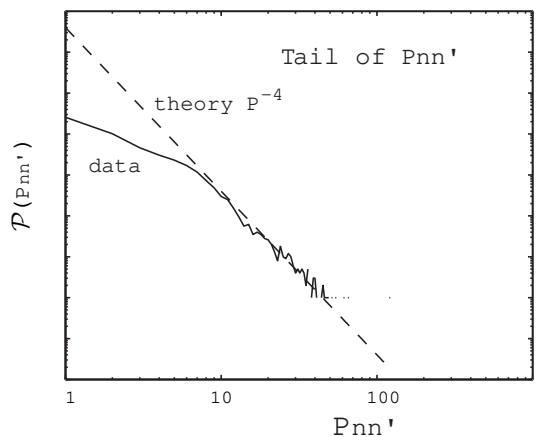,width=7cm}}
\vskip 0.05in
\caption{The tail region of the plot of the $P_{nn'}$ 
distribution  shows
good agreement with the predicted quartic power law,
for a $2\pi \times 2\pi$
billiard, $f = 0.23$, 226 channels.  
}
\label{nnnt}
\end{figure}

Finally we consider the tails of the intensity and transport measures.
From Eq.~\ref{ppntail} we expect a cubic fall off in the 
tail of the $P_{n\xi}$ intensity distribution:
${\cal P}(P_{n\xi} =x) \sim { 1 / f x^3 }$ for $x \gg  1/f$.
In Fig.~\ref{ppnt} we display the predicted and numerical
results, showing good agreement between the two.
This behavior is controlled
by the near-bouncing ball dynamics. (We do not discuss again the
behavior of the intensity distribution intermediate between
the head and tail; 
in Fig.~\ref{kickfig6} we saw already that the entire distribution
is well-predicted classically.)

The tail of the transport distribution measure ${\cal P}(P_{nn'})$ is 
given by Eq.~\ref{quar},
${\cal P}(P_{nn'} = x) \sim { 1 /f^2 x^4}$.
Fig.~\ref{nnnt} again demonstrates  very good agreement with this estimate.
Notice that the RMT prediction is $P_{nn'}=1$ for all channels
$n \ne n'$.

\section{Simple Unitary Matrix Model}
\label{matrix}

The previous examples corresponded to physical systems, or maps, which have a 
direct basis in dynamics.  Above, we have  compared the results
for such dynamical systems to random matrix theory.  However
there is a variant of the usual random matrix theory, i.e. a modified
random matrix ensemble, which retains some of the 
gross characteristics of our dynamical systems, while remaining 
free of any real dynamics.  The main idea is to retain the tendency to 
scatter back into the same channel (diagonal dominance)
while making that portion of the 
amplitude which does scatter do so randomly. This gives rise to Lorentzian
envelopes as in the dynamical systems, but not with the
near-bouncing ball effects,  which strongly skewed the tails of the
intensity, IPR, and transport
distributions discussed above.  The S-matrix for the collision
off the Sinai obstruction in the corridor is the key element in the
theory of the eigenstates presented above. The S-matrix for this
process is unitary and symmetric.  The fully random matrix
ensemble corresponding to a symmetric  S-matrix 
is Dyson's circular orthogonal ensemble, the COE\cite{mehta}.
However we wish to modify this to a diagonally
dominated symmetric unitary matrix $U$ which includes the 
effect of random off-diagonal 
coupling of variable strength.  The
random component schematically represents the scattering off a small
object, with none of the subsequent dynamical correlations built in.
We take   the form 
\begin{equation} 
U = \exp[i(D + R)]
\end{equation} 
where $D$ is a diagonal matrix with randomly chosen quasienergies
on the interval $[0, 2\pi)$, $R$ is a GOE random matrix.
 Ensembles similar to this have
been used previously to model spectral statistics intermediate
between Wigner-Dyson and Poisson~\cite{interm}.

For large $ \gamma$ we approach the COE limit; one iteration of the 
$U$ matrix on a starting vector will then decorrelate it completely.  That 
is, the self-overlap becomes
\begin{equation} 
{\vert \langle n\vert U \vert n\rangle \vert^2 \over 
\vert \langle n\vert n\rangle \vert^2}\sim {1\over N} \,.
\end{equation} 
This one step decay corresponds to a pseudoenergy
uncertainty of $2 \pi$, which is 
just that required to give a uniform spectral density on $[0, 2\pi)$.
For smaller $ \gamma$ we have slower decay, going as 
\begin{equation} 
\label{decay}
{\vert \langle n\vert U^m \vert n\rangle \vert^2 \over 
\vert \langle n\vert n\rangle \vert^2}\sim \exp[-2\gamma m],
\end{equation} 
which leads to a Lorentzian lineshape as in Eq.~\ref{snaive}.

The measures of distributions, tails, etc. above can be defined
for the Lorentzian envelope (modified COE) model as well.  The 
situation here is less rich, since almost all states decay with 
approximately the same rate, unlike the special channels (angles)
in the Sinai 
models  which dictate anomalously slow decay. The IPR is anomalous,
up by a factor of $\sim \gamma^{-1}$ from the RMT prediction
due to the Lorentzian LDOS envelopes. Transport is similarly
anomalous, with each channel coupled to only a fraction $\sim \gamma$
of all the other channels at long times ($Q \sim 1/\gamma$).

Due to the near-bouncing-ball orbits, the average IPR in the Sinai systems
revealed a localization increasing as $\log N/f$ as the
classical limit $N \to \infty$ was taken. For the unitary matrix
model, we have only a 
single decay rate $\gamma$ (not the distribution caused by the
near bouncing ball orbits);
the resulting average IPR is therefore
predicted to be independent of $N$.  
The decay rate $\gamma$ is given by the variance of the 
matrix $R$ through the Golden Rule,
\begin{equation} 
\gamma = 2\pi \langle R^2 \rangle \rho
\end{equation} 
with $\rho = N/2\pi$.
In the limit of small $\gamma$, the IPR should go as $3\cdot 2/\gamma$,
where the 
factor of three is  the COE fluctuation factor. The table 
below shows the mean
IPR's averaged over all the basis 
states for two values each of $N$ and $\gamma$ using the modified COE.
Excellent agreement is seen between  the predicted and found IPR's.
\vskip 0.1in
\centerline{
\begin{tabular}{|c|c|c|c|}
\hline
\multicolumn{4}{|c|}{\rule[-3mm]{0mm}{8mm}
  \bf Table of IPR for modified COE}\cr 
  \hline 
  $\gamma \vert N$ & $350$ & $450$ & {\rm IPR = $6/\gamma$} \cr
  \hline \hline
{0.084}&{62.6}&{65.9}&{IPR = 71.4} \cr\hline
{0.188}&{32.9}&{34.7}&{IPR = 31.9} \cr\hline
\end{tabular}}
\vskip 0.1in
The long power-law tails in the intensity, IPR, and transport efficiency
distributions, present in the real dynamical Sinai system, are
similarly absent in the modified COE model.

\section{Conclusion}

Random matrix ensembles possess eigenstates which are maximally random,
consistent with the symmetry constraints which govern the 
particular ensemble. The properties of such eigenstates form the basis of
much work in quantum chaos theory,
and more importantly the basis of much theory
of nuclei, molecules, and especially mesoscopic devices.  Is random matrix
theory the limit to which real
classically chaotic systems adhere as $\hbar \to 0$?
Definitely not.

The SZCdV theory  predicts only coarse grained ergodicity
of individual eigenstates in the $\hbar \to 0$ limit, which is much weaker than
the requirements of random matrix ensembles.  This gap, between 
random matrix ensembles on the one hand and SZCdV on the other,
leaves open many questions about the true nature of eigenstates of classically
chaotic systems in the $\hbar \to 0$ limit.  We have been engaged
for some time in the exploration of these questions, which address the 
fluctuations of eigenstates on scales that shrink
as some positive fractional power of 
Planck's constant (or, more physically, as some negative fractional
power of the energy).
Since such scales become infinitesimal as   $\hbar \to 0$,
SZCdV has little to say about them. Yet they may contain infinitely many 
wavelengths in this limit.
The earliest work in this area is scar theory\cite{scars,wqe,is,unst},
which showed  that the effects of the least unstable
periodic orbits survive the $\hbar \to 0$ limit.
However this is only one possible
type of a non-RMT ``anomaly'' in classically chaotic systems. 

Our first investigation beyond scar theory, using
the so called tilted-wall billiard, examined a very slowly
ergodic classical system with the expectation that its eigenstates would be 
maximally likely to show non-RMT behavior. 
Indeed the eigenstates  did show increasing localization
on small scales as  $\hbar \to 0$,
while still of course obeying the SZCdV ergodic theorem.

In the present study we have switched to the traditional
paradigm of classical chaos, namely
the Sinai billiard and some close cousins.
We have been able to show that the eigenstates 
are ever more strongly localized in a certain basis as $\hbar \to 0$.
The basis used is 
not extraordinary: essentially it is 
the  usual plane waves (momentum space) of scattering theory.   
We showed that the mean inverse participation ratio in the
Sinai-like systems diverges logarithmically with increasing energy
(or decreasing $\hbar$), implying  that wavefunctions are becoming less  
ergodic at the single-channel scale  
as the classical limit is approached.
The situation
here is more remarkable than in the tilted billiard~\cite{wqe}, since
in Sinai systems the Lyapunov exponent is positive and classical
correlations decay exponentially.
A major conclusion of this work is that the logarithmically
increasing mean IPR in
Eq.~\ref{logmean} is not due to the bouncing ball states  
but instead to the ``near-bouncing ball" channels, whose decay time
is large compared to the typical decay time $1/f$ but still small compared
to the Heisenberg time $N$ at which individual eigenstates are resolved.

Another key point is that short time quantum 
dynamics and correlation functions have an irreversible effect on the 
localization properties of the eigenstates, as in the case of scar theory.  
 
Undoubtedly there are many more  non-RMT effects 
in eigenstates yet be uncovered in other systems, 
including some that could affect important
physical properties.  

\section{Acknowledgments}
This research was supported by the National Science Foundation under
Grant CHE-9610501.

\end{document}